\documentclass[%
 reprint,
superscriptaddress,
nofootinbib,
eqsecnum,
longbibliography,
 amsmath,amssymb,
 aps,
]{revtex4-1}

\usepackage[dvipsnames]{xcolor}
\usepackage{graphicx}
\usepackage{dcolumn}
\usepackage{bm}
\usepackage[colorlinks]{hyperref}
\usepackage{cleveref}
\usepackage{tabularx}
\usepackage{xspace}
\usepackage[normalem]{ulem}
\usepackage[utf8]{inputenc}

\usepackage[inline]{enumitem}

\newcommand\myshade{80}
\colorlet{mylinkcolor}{ForestGreen}
\colorlet{mycitecolor}{Red}
\colorlet{myurlcolor}{violet}
\usepackage[dvipsnames]{xcolor}

\hypersetup{
  linkcolor  = mylinkcolor!\myshade!black,
  citecolor  = mycitecolor!\myshade!black,
  urlcolor   = myurlcolor!\myshade!black,
  colorlinks = true
}

\newcommand{\rmsp}{\mathrm{sp}} 
\newcommand{\rmDM}{\mathrm{DM}} 
\newcommand{\rmIn}{\mathrm{in}} 
\newcommand{\rmenc}{\mathrm{enc}} 
\newcommand{\rmsh}{\mathrm{sh}} 
\newcommand{\rmGW}{\mathrm{GW}} 
\newcommand{\rmDF}{\mathrm{DF}} 
\newcommand{\rmisco}{\mathrm{ISCO}} 
\newcommand{\rmi}{\mathrm{i}} 
\newcommand{\rmf}{\mathrm{f}} 
\newcommand{\calE}{\mathcal{E}}

\newcommand{\EdaEtAl}{Eda \textit{et al.}\xspace}

\newcommand{\GRAPPA}{%
Gravitation Astroparticle Physics Amsterdam (GRAPPA),\\
Institute for Theoretical Physics Amsterdam
and Delta Institute for Theoretical Physics,\\
University of Amsterdam, Science Park 904, 1098 XH Amsterdam, The Netherlands}

\newcommand{\UVA}{%
Department of Physics, University of Virginia, P.O.~Box 400714, Charlottesville, VA 22904-4714, USA}

\newcommand{\Radboud}{%
Department of Astrophysics, Faculty of Science, Radboud University Nijmegen, P.O.~Box 9010, 6500 GL Nijmegen, The Netherlands}

\newcommand{\IFT}{
Instituto de F\'isica Te\'orica UAM-CSIC, \\
Campus de Cantoblanco, E-28049 Madrid, Spain}

\begin{document}

\preprint{}

\title{Detecting dark matter around black holes with gravitational waves:\\ Effects of dark-matter dynamics on the gravitational waveform}

\author{Bradley J. Kavanagh}%
 \email{b.j.kavanagh@uva.nl}
\affiliation{\GRAPPA}%

\author{David A. Nichols}%
\email{david.nichols@virginia.edu}
\affiliation{\UVA}
\affiliation{\GRAPPA}
\affiliation{\Radboud}%

\author{Gianfranco Bertone}
\email{g.bertone@uva.nl}
\affiliation{\GRAPPA}

\author{Daniele Gaggero}
\email{daniele.gaggero@uam.es}
\affiliation{\IFT}

\date{\today}

\begin{abstract}
A dark matter overdensity around a black hole may significantly alter the dynamics of the black hole's merger with another compact object.
We consider here intermediate mass-ratio inspirals of stellar-mass compact objects with intermediate-mass black holes ``dressed" with dark matter. 
We first demonstrate that previous estimates based on a fixed dark-matter dress are unphysical for a range of binaries and dark-matter distributions by showing that the total energy dissipated by the compact object through dynamical friction, as it inspirals through the dense dark matter environment towards the black hole, is larger than the gravitational binding energy of the dark-matter dress itself.
We then introduce a new formalism that allows us to self-consistently follow the evolution of the dark-matter dress due to its gravitational interaction with the binary. We show that the dephasing of the gravitational waveform induced by dark matter is smaller than previously thought, but is still potentially detectable with the LISA space interferometer.
The gravitational waves from such binaries could provide powerful diagnostics of the particle nature of dark matter.
\end{abstract}

\maketitle


\section{Introduction}
The direct detection of gravitational waves (GWs)~\cite{Abbott:2016blz,TheLIGOScientific:2017qsa,GBM:2017lvd,LIGOScientific:2018mvr} has opened up new opportunities for fundamental physics. Present and upcoming experiments such as LIGO/Virgo \cite{LIGOScientific:2019vkc,TheVirgo:2014hva}, KAGRA \cite{Akutsu:2019rba}, LISA \cite{AmaroSeoane:2012km,2017arXiv170200786A}, Einstein Telescope \cite{Sathyaprakash:2012jk} and pulsar timing arrays
\cite{2010CQGra..27h4013H,2013CQGra..30v4009K,Hobbs:2013aka,2009arXiv0909.1058J} will soon shed light on a variety of problems at the intersection between gravitational waves, black holes and fundamental physics~\cite{Barack:2018yly}, and, in particular, on the distribution and nature of dark matter (DM) \cite{Bertone:2018xtm,Bertone:2019irm}.

Here, we focus on the prospects for detecting and characterizing cold dark-matter overdensities around black holes (BHs) using gravitational waves. 
If dark matter is made of cold collisionless particles, the adiabatic growth of black holes may induce the formation of large overdensities (often referred to as ``spikes") around supermassive \cite{Quinlan:1994ed,Gondolo:1999ef,Ullio:2001fb} and intermediate-mass \cite{Bertone:2005hw,Zhao:2005zr,Bertone:2006nq} astrophysical black holes, as well as around primordial black holes \cite{Kohri:2014lza,Eroshenko:2016yve,Boucenna:2017ghj,Hertzberg:2019exb}. It is in principle possible to detect and characterize DM overdensities around black holes by measuring their impact on the gravitational waveform as BHs merge with other compact objects \cite{Eda:2013gg,Eda:2014kra,Macedo:2013qea,Barausse:2014tra,Barausse:2014pra,Yue:2017iwc,Yue:2019ndw,Hannuksela:2019vip,Cardoso:2019rou}.

\begin{figure}[tb!]
    \begin{center}
    \includegraphics[width=0.8\columnwidth,scale=0.8]{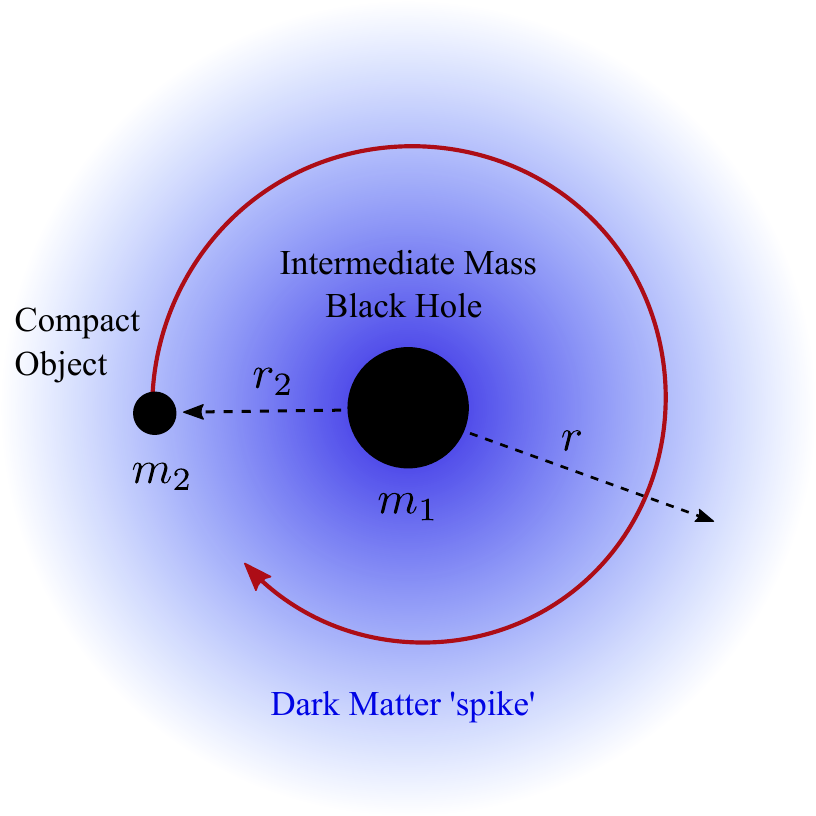}
    \end{center}
    \caption{\textbf{Intermediate mass-ratio inspiral (IMRI) system with a dark matter ``spike.''} A central intermediate-mass black hole (IMBH) of mass $m_1$ is orbited by a lighter compact object $m_2 < m_1$ at an orbital radius $r_2$. The IMBH is also surrounded by a ``spike'' of dark matter with density profile $\rho_\mathrm{DM}(r)$.}
    \label{fig:IMRI}
\end{figure}

In this paper, we revise previous calculations of the orbital evolution of and gravitational waveforms from intermediate mass-ratio inspirals (IMRIs) around ``dressed'' black holes, as illustrated in Fig.~\ref{fig:IMRI}. In such a system, a stellar-mass compact object (black hole or neutron star) inspirals towards an intermediate-mass black hole (IMBH) with mass $10^{3}$ -- $10^{5} \, M_\odot$. The presence of DM exerts a dynamical friction force~\cite{Chandrasekhar1943a,Chandrasekhar1943b,Chandrasekhar1943c} on the compact object, causing it to inspiral more rapidly. The resulting gravitational waveform accumulates phase at a different rate compared to the vacuum case (in the absence of DM). This ``dephasing'' effect should be detectable with future GW observatories, but accurate waveform modeling is required to extract the signal and perform parameter estimation~\cite{Arnaud:2006gm,Babak:2009cj,LISA_datachallenge}.

We begin by exploring energy conservation in these systems, and we show that the work done by dynamical friction is typically comparable to (and in some cases much larger than) the total binding energy available in the DM spike. This means that previous calculations of the de-phasing signal, which assumed a non-evolving dark-matter density profile, do not conserve energy and therefore substantially overestimate the size of the effect. 
In order to develop a self-consistent description of such systems, we first present $N$-body simulations which allow us to accurately model dynamical friction and the scattering of DM particles with the compact object. In particular, this allows us to understand where the energy lost by the compact object is injected in the DM cloud. We then devise a prescription for evolving the phase space distribution of DM as energy is injected during the inspiral. 

We self-consistently follow the evolution of the binary and the DM spike, and we robustly estimate the dephasing of the gravitational waveform with respect to both the vacuum inspiral, and to the unphysical case of a static DM halo. We demonstrate that the dephasing of the gravitational waveform induced by dark matter is smaller than previously assumed, but is still potentially detectable by the LISA mission, which will have a peak sensitivity at frequencies between $10^{-3}$ and $10^{-2}$~Hz \cite{Cornish:2018dyw}.
It could thus provide a powerful diagnostic of the particle nature of dark matter.

The paper is organized as follows: in Sec.~\ref{sec:EnergyBalance}, we demonstrate that the standard approach to the dephasing signal induced by DM minispikes is likely to violate energy conservation; in Sec.~\ref{sec:Nbody}, we present $N$-body simulations to validate our model for dynamical friction; in Sec.~\ref{sec:HaloFeedback}, we present our prescription for evolving the phase space distribution of DM; in Sec.~\ref{sec:Results}, we use this prescription to follow the evolution of the binary and the DM spike self-consistently; finally, in Sec.~\ref{sec:Discussion}, we discuss some caveats of this work and possible implications for the detection of such a DM spike in intermediate mass-ratio inspirals in the future.
We conclude in Sec.~\ref{sec:Conclusions}, and we have several supplementary results in four appendices.

\section{Energy balance considerations for static dark matter halos}
\label{sec:EnergyBalance}

In this section, we describe the evolution of a system composed of a central IMBH with a surrounding DM spike and a lighter compact object (e.g.\ a neutron star) orbiting around the IMBH and through its DM cloud. This is illustrated in Fig.~\ref{fig:IMRI}.  We model the evolution of this system using Newtonian gravity, and we include dissipative effects arising from dynamical friction and gravitational radiation. 
Following \EdaEtAl~\cite{Eda:2013gg,Eda:2014kra}, we neglect any feedback on the DM halo in this section, and we consider only circular orbits.

\subsection{Notation for IMBH system and DM distribution}
\label{subsec:notation}

We first define several notions of masses for the binary and the DM distribution.
We will denote the mass of the IMBH by $m_1$ and the mass of the small compact object by $m_2$.
Other definitions of masses we will need are  $M=m_1 + m_2$, the total mass; $q=m_2/m_1 \leq 1$, the mass ratio; $\mu = m_1 m_2 / M$, the reduced mass; and $\mathcal M_c = \mu^{3/5} M^{2/5}$, the chirp mass.

We assume that the IMBH is surrounded by a DM spike, formed as the adiabatic growth of the black hole enhances the central density of the host halo \cite{Gondolo:1999ef,Bertone:2005hw,Sadeghian:2013laa,Ferrer:2017xwm,PhysRevD.101.024029}. The dark-matter distribution will be given by 
\begin{equation} \label{eq:rhoDM}
\rho_\rmDM(r) = \left\{
\begin{array}{ll}
  \rho_\rmsp \left(\frac{r_\rmsp}{r}\right)^{\gamma_\rmsp}   &  r_\rmIn \leq r \leq r_\rmsp\\
   0.  & r < r_\rmIn
\end{array}
\right. \, ,
\end{equation}
where $r$ is the distance from the center of the IMBH.
We define the inner radius of the spike to be $r_\rmIn = 4 G m_1 / c^2$ following the results in~\cite{Sadeghian:2013laa}.
We will not treat the DM distribution at distances $r > r_\rmsp$.
We also will not treat $r_\rmsp$ as a free parameter, but as determined by $m_1$, $\rho_\rmsp$ and $\gamma_\rmsp$ via
\begin{equation} \label{eq:rsp}
    r_\rmsp \approx \left[ \frac{(3-\gamma_\rmsp)0.2^{3-\gamma_\rmsp}m_1}{2\pi \rho_\rmsp}\right]^{1/3} \, .
\end{equation}
This assumes that $r_\rmsp \approx 0.2 r_{\rm h}$, where $r_{\rm h}$ is defined from
\begin{equation}
    \int_{r_\rmIn}^{r_{\rm h}} \rho_\rmDM(r) 4\pi r^2 \,\mathrm{d}r = 2 m_1 \, ,
\end{equation}
as in~\cite{Eda:2014kra}. We can now compute the DM mass within a distance $r$.
The result is
\begin{equation} \label{eq:menc}
    m_\rmenc(r) = \left\{ \begin{array}{ll} m_\rmDM(r) - m_\rmDM(r_\rmIn) &  r_\rmIn \leq r \leq r_\rmsp\\
   0.  & r < r_\rmIn
\end{array}  \right. \, ,
\end{equation}
where
\begin{equation}
    m_\rmDM(r) = \frac{4\pi \rho_\rmsp r_\rmsp^{\gamma_\rmsp}}{3-\gamma_\rmsp}r^{3-\gamma_\rmsp} \, .
\end{equation}

With this notation set, we can now more easily discuss issues related to energy balance.

\subsection{Gravitational potential energy of the DM distribution}
\label{subsec:Potential}

To compute the total potential energy in the distribution of DM, we determine the amount of work required to assemble the distribution of DM by adding successive spherical shells of DM of increasing radius $r$, until the final distribution $\rho_\rmDM(r)$ is constructed around the BH.
We denote the potential energy of each shell of DM of radius $r$ by $\mathrm{d}U_\rmsh(r)$.
It is given by
\begin{equation}
    \mathrm{d}U_\rmsh(r) = -\frac{G [m_1 + m_\rmenc(r)]}{r} [4\pi r^2 \rho_\rmDM(r) \,\mathrm{d}r] \, .
\end{equation}
After some algebra, we can instead write it as
\begin{equation} \label{eq:Ushell}
    \mathrm{d}U_\rmsh(r) = -\frac{G [m_1 + m_\rmenc(r)]m_\rmDM(r) (3-\gamma_\rmsp) \,\mathrm{d}r}{r^2}  \, .
\end{equation}

Integrating Eq.~\eqref{eq:Ushell} between the inner radius $r_\rmIn$ and a given radius $r$, we arrive at the total potential energy in the distribution of DM between the radii $r_\rmIn$ and $r$.
When $\gamma_\rmsp \neq 2$ or $\gamma_\rmsp \neq 5/2$, the result is
\begin{equation} \label{eq:DeltaUDM}
\begin{split}
    \Delta U_\rmDM (r) = & -\frac{G m_\rmDM(r) (3-\gamma_\rmsp)}{r} \\
    & \times \left[\frac{m_1 - m_\rmDM(r_\rmIn)}{2-\gamma_\rmsp} + \frac{m_\rmDM(r)}{5-2\gamma_\rmsp} \right]
    - U_\rmIn \, ,
\end{split}
\end{equation}
where the constant $U_\rmIn$ is given by
\begin{equation}
    U_\rmIn = -\frac{G m_\rmDM(r_\rmIn) (3-\gamma_\rmsp)}{r_\rmIn {(2-\gamma_\rmsp)}}  \\
    \left[m_1 - \frac{m_\rmDM(r_\rmIn) (3 - \gamma_\rmsp)} {5-2\gamma_\rmsp} \right]\, .
\end{equation}
The total potential energy of the DM spike can be obtained by evaluating Eq.~\eqref{eq:DeltaUDM} at $r=r_\rmsp$.

Note that we are ignoring the effect of the gravitational potential of the small compact object on the binding energy.
This will generally lead to relative errors of order $q$, which will be small for the systems we are considering.

\subsection{Orbital energy and energy dissipation through GWs and DF}
\label{subsec:EnergyDissipate}

Next, we will summarize how we compute the orbital energy and the
dissipation of orbital energy through gravitational waves and
dynamical friction.
Our formalism is similar to that presented in \EdaEtAl~\cite{Eda:2013gg,Eda:2014kra}. 
Since the system we are considering is characterized by a small mass ratio between the IMBH and the orbiting compact object ($q\ll 1$), we will adopt the approximation $\mu \simeq m_2$
(the errors in this approximation are of order $q$).
This assumes that the barycenter position is equal to the IMBH position.
Similarly, assuming $M=m_1$ leads to errors of order $q$.
We discuss the impact of this approximation in more detail in
Sec.~\ref{sec:Discussion}.
We will also work with circular orbits, and we will ignore the correction to the Keplerian frequency arising from the distribution of DM (which will be a percent-level effect for most of the binaries we study in this paper).
In this approximation, the orbital energy reduces to the familiar expression
\begin{equation}
\label{eq:Eorbit}
    E_{\rm orb} = - \frac{G m_1 m_2}{2r_2} \, .
\end{equation}

Since the lighter object moves within the DM mini-spike and experiences gravitational interactions with the DM particles, it loses energy via {\it dynamical friction} (DF)  \cite{Chandrasekhar1943a,Chandrasekhar1943b,Chandrasekhar1943c}. 
In addition, the orbital energy changes through the emission of gravitational waves. 
The timescale over which energy is dissipated through these processes is slow compared to the orbital timescale for most of the evolution of the system.
Thus, we will treat the dissipation as an adiabatic process slowly moving the compact object on a given circular orbit to another circular orbit with a slightly smaller radius (i.e.\ a quasi-circular inspiral).
In this process, energy balance is satisfied, in the sense that 
\begin{equation}
\frac{{\rm d}E_{\rm orb}}{{\rm d}t} \,=\, - \frac{{\rm d}E_{\rm GW}}{{\rm d}t} - \frac{{\rm d}E_{\rm DF}}{{\rm d}t} \, .
\end{equation}
Gravitational-wave energy losses (for circular orbits in the quadrupole approximation) are given by
\begin{equation}
\label{eq:GWdissipation}
    \frac{{\rm d}E_{\rm GW}}{{\rm d}t} = \frac{32 G^4 M (m_1 m_2)^2}{5 (c r_2)^5} \, . 
\end{equation}
Dynamical friction losses are given by
\begin{equation}
\label{eq:DFdissipation}
   \frac{{\rm d}E_{\rm DF}}{{\rm d}t} = 4\pi (G m_2)^2 \rho_\rmDM(r_2) \,\xi(v)\, v^{-1} \log\Lambda \, . 
\end{equation}
The term $\xi(v)$ denotes the fraction of DM particles moving more slowly than the orbital speed.\footnote{This term has typically been neglected in previous studies of DM dephasing \cite{Eda:2013gg,Eda:2014kra}. For the isotropic spike profile with $\gamma_\mathrm{sp} = 7/3$ around an IMBH of mass $10^3\,M_\odot$, we find $\xi(v) \approx 0.58$, independent of radius. We set $\xi=1$ in the analytic analysis of this section, though as we will see in Sec.~\ref{sec:Nbody}, it will be necessary to include it later to obtain an accurate description of the dynamics.}

In Eq.~\eqref{eq:DFdissipation}, $\log\Lambda$ is the usual notation for the Coulomb logarithm, defined in general as \cite[App.~L]{BinneyAndTremaine}:
\begin{equation}
\label{eq:coulombLog}
\Lambda=\sqrt{\frac{b_{\max }^{2}+b_{90}^{2}}{b_{\min }^{2}+b_{90}^{2}}}\,,
\end{equation}
where $b_{\mathrm{min}}$ and $b_{\mathrm{max}}$ are the minimum and maximum impact parameters for which the two-body encounters that contribute to the phenomenon can be considered effective. 
Moreover, $b_{90}$ is the impact parameter which produces a $90^\circ$ deflection of the DM particle:
\begin{equation}
\label{eq:b90}
b_{90} = \frac{G m_2}{v_0^2} \approx \frac{m_2}{m_1} r_2 = q \,r_2\,,
\end{equation}
with $v_0$ the orbital speed of the compact object. We fix $\Lambda=\sqrt{m_{1} / m_{2}}$, as we discuss in more detail in Sec.~\ref{sec:Nbody}.

It will be convenient to write these losses as a function of $r_2$ for circular orbits by using the relationship that $v=\sqrt{GM/r_2}$.
Using the chain rule and Eqs.~\eqref{eq:Eorbit}, \eqref{eq:GWdissipation}, and \eqref{eq:DFdissipation}, we can also write an explicit expression for the time evolution of the small compact object's separation:
\begin{equation} \label{eq:r_dot}
\dot{r}_2 \,=\, - \frac{64\, G^3\, M \, m_1\, m_2}{5\, c^5\, (r_2)^3} - \frac{8 \pi\, G^{1/2}\, m_2 \, \rho_\rmsp \, \xi\,\log\Lambda \, r_\rmsp^{\gamma_\rmsp}}{\sqrt{M} m_1 \,\, r_2^{\gamma_\rmsp - 5/2}}  \, .
\end{equation}
As the small compact object inspirals between circular orbits with two radii $r_\rmi$ and $r_\rmf$ (with $r_\rmi > r_\rmf$), some fraction of the orbital energy will be carried away by GWs, and some fraction will be dissipated through dynamical friction.
We write this as
\begin{equation}
    \Delta E_{\rm orbit} = \Delta E_\rmDF + \Delta E_\rmGW \ .
\end{equation}
While the energy dissipated by GW emission is expected to have a negligible effect on the distribution of DM, the energy dissipated through DF will go directly into increasing the energy of the particles in the DM distribution.

Because the DM spike has a finite amount of potential energy, $\Delta U_\rmDM(r_\rmsp)$, it is important to check that the energy dissipated through dynamical friction, $\Delta E_\rmDF$, is not comparable to (or in excess of) $\Delta U_\rmDM(r_\rmsp)$.
If they are comparable, then this would imply that enough energy is dissipated through DF to alter significantly the distribution of DM (and perhaps even to unbind all the DM from the gravitational potential of the IMBH).
Even when the ratio $\Delta E_\rmDF / \Delta U_\rmDM(r_\rmsp)$ is comparable to but less than one, then it is generally not a good approximation that the distribution of DM would remain invariant during the inspiral of the small compact object.

Thus, it is important to compute the total energy dissipated through dynamical friction $\Delta E_\rmDF$ during the inspiral.
This can be found by integrating Eq.~\eqref{eq:DFdissipation} between two given times, or more conveniently, integrating the following expression between two radii, $r_\rmi$ and $r_\rmf$ describing two circular orbits:
\begin{equation} \label{eq:DeltaEDFr}
    \Delta E_\rmDF(r_\rmi,r_\rmf) = -\int_{r_\rmi}^{r_\rmf} \frac{{\rm d}E_{\rm DF}}{{\rm d}t} \left(\frac{{\rm d}r_2}{{\rm d}t} \right)^{-1} \,\mathrm{d}r_2 \, . 
\end{equation}
In Eq.~\eqref{eq:DeltaEDFr}, the radial evolution equation is defined in~\eqref{eq:r_dot}, and the dynamical friction energy loss is defined in~\eqref{eq:DFdissipation}. 
After some algebra, the integral in~\eqref{eq:DeltaEDFr} can be expressed as 
\begin{equation} \label{eq:DeltaEDFrInt}
    \Delta E_\rmDF = - \frac{G m_1 m_2}2 \int_{r_\rmi}^{r_\rmf} \frac{\mathrm{d}r_2}{(r_2)^2(1+c_r r^{-11/2 + \gamma_\rmsp})} \, ,
\end{equation}
where 
\begin{equation}
    c_r = \frac{8 G^{5/2} M^{3/2} (m_1)^2}{5\pi c^5 \rho_\rmsp r_\rmsp^{\gamma_\rmsp} \xi \log \Lambda} 
\end{equation}
and where for simplicity, in this section, we assume $\xi=1$.
The integral~\eqref{eq:DeltaEDFrInt} can be evaluated in terms of hypergeometric functions as follows:
\begin{equation} \label{eq:DeltaEDFhyp}
\begin{split}
    \Delta E_\rmDF & = \left[ \frac{G m_1 m_2}{2r_2} \right. \\
    & \left.\times {}_2F_1\left(1, \frac{2}{11-2\gamma_\rmsp}, \frac{13-2\gamma_\rmsp}{11-2\gamma_\rmsp}; -c_r r_2^{-11/2+\gamma_\rmsp} \right)\right]^{r_\rmi}_{r_\rmf} .
\end{split}
\end{equation}
This expression has an interesting form: because the hypergeometric function is a number in the range $(0,1)$ for positive $r_2$, then~\eqref{eq:DeltaEDFhyp} represents the difference between two fractions of the energy of two circular orbits at two radii.

Thus, with Eqs.~\eqref{eq:DeltaUDM} and~\eqref{eq:DeltaEDFhyp}, we can compute ratios of energy dissipated by dynamical friction to binding energy in the DM distribution surrounding the IMBH.

\subsection{Ratio of energy dissipated to binding energy}
\label{subsec:EnergyRatio}

In \EdaEtAl, the system investigated in greatest detail is a binary in which the IMBH has mass $m_1=10^3 M_\odot$ and the small compact object has mass $m_2 = 1\,M_\odot$.
The DM spike is characterized by a density normalization $\rho_\rmsp = 226 M_\odot/\mathrm{pc}^3$ and a power law $\gamma_\rmsp = 7/3$ (the corresponding value of $r_\rmsp$ is 0.54 pc). The slope $\gamma_\mathrm{sp} = 7/3$ is expected to develop in the center of a halo with an initial profile scaling as $\rho \sim r^{-1}$, such as an NFW profile~\cite{Gondolo:1999ef}.
\EdaEtAl observe that during the last five years as the small compact object inspirals towards the IMBH before merging, the effect of dynamical friction can significantly change the rate of inspiral.
The large change in the inspiral occurs because a significant amount of energy is dissipated through dynamical friction (and thus must be balanced by increasing the kinetic energy of the DM particles in the halo).

\begin{figure}[tb!]
    \centering
    \includegraphics[width=\columnwidth]{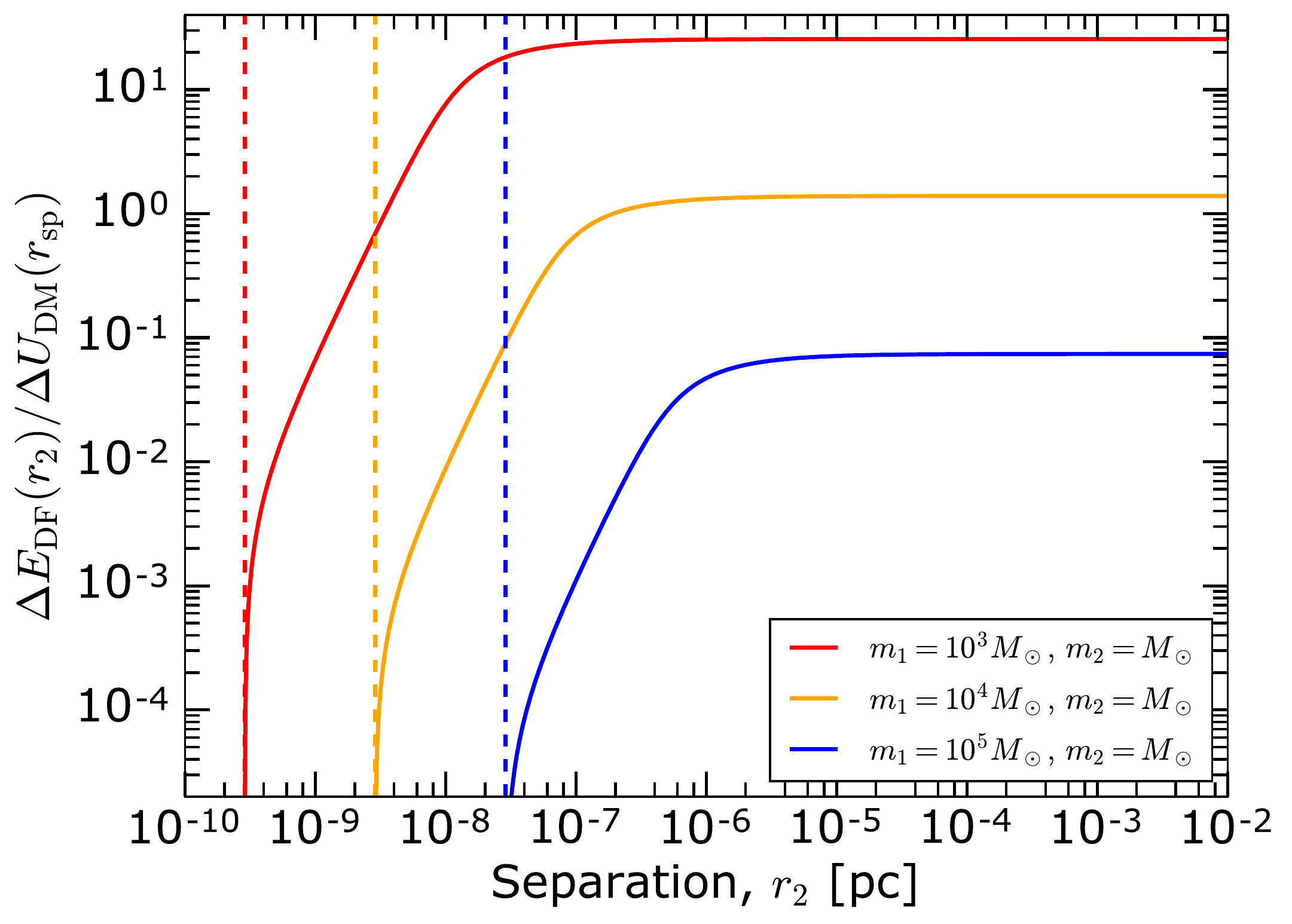}
    \caption{\textbf{Ratio of energy radiated through dynamical friction to binding energy of the DM spike versus separation}.
    The solid curves (red, orange, and blue) correspond to
    three different mass ratios for three binaries ($q = 10^{-3}$, $10^{-4}$, and $10^{-5}$, respectively).
    The dashed vertical lines correspond to the ISCO radii for
    the three binaries. Here, we assume $\rho_\rmsp = 226 M_\odot/\mathrm{pc}^3$ and $\gamma_\rmsp = 7/3$ for the DM spike.}
    \label{fig:EDFbyUDMcurves}
\end{figure}

In Fig.~\ref{fig:EDFbyUDMcurves}, we define the energy 
dissipated between a separation $r_2$ and $r_\rmisco$ by
\begin{equation}
    \Delta E_\rmDF(r_2) \equiv \Delta E_\rmDF(r_2,r_\rmisco) \, ,
\end{equation}
and we plot the ratio of this energy to the total binding energy
of the DM spike, $\Delta U_\rmDM(r_\rmsp)$ as a function of
separation $r_2$.
The three solid curves in red, orange, and blue correspond to
binaries with mass ratios $q=10^{-3}$, $10^{-4}$, and $10^{-5}$.
In all three cases, the following three parameters are the same:
$\rho_\rmsp = 226 M_\odot/\mathrm{pc}^3$, $\gamma_\rmsp = 7/3$, 
and $m_2 = M_\odot$.
The vertical dashed lines show the positions of the ISCO radius
for the three cases.

The figure highlights a few important points.
First, for all three mass ratios shown, the quantity 
$E_\rmDF(r_2)/\Delta U_\rmDM(r_\rmsp)$ grows rapidly with $r_2$
out to a few hundred ISCO radii, and then it plateaus to a
nearly constant value at larger separations.
Because the ratio $E_\rmDF(r_2)/\Delta U_\rmDM(r_\rmsp)$
is nearly constant over a large range of radii, we will use
the number $E_\rmDF(r_\rmsp)/\Delta U_\rmDM(r_\rmsp)$ as a 
figure of reference for the characteristic fraction of energy
dissipated to the binding energy of the halo.
Second, these curves show that for more equal mass ratios, 
the mismatch between the amount of binding energy in the halo 
and the amount of energy dissipated through dynamical friction
becomes worse.\footnote{Because $m_\rmDM(r)$ satisfies the property $m_\rmDM(r_\rmsp) \sim m_1$, then it is not too difficult to see that the ratio $\Delta E_\rmDF(r_\rmsp) / \Delta U_\rmDM(r_\rmsp)$ will scale linearly with the mass ratio.}
Using $E_\rmDF(r_2)/\Delta U_\rmDM(r_\rmsp) \approx 1$ as a 
rough figure of merit, the $10^{-3}$ mass ratio system poorly
preserves energy balance, the $10^{-4}$ system roughly satisfies
energy balance, and the $10^{-5}$ system does not run into issues
with energy balance.

\begin{figure*}[htb]
    \centering
    \includegraphics[width=\textwidth]{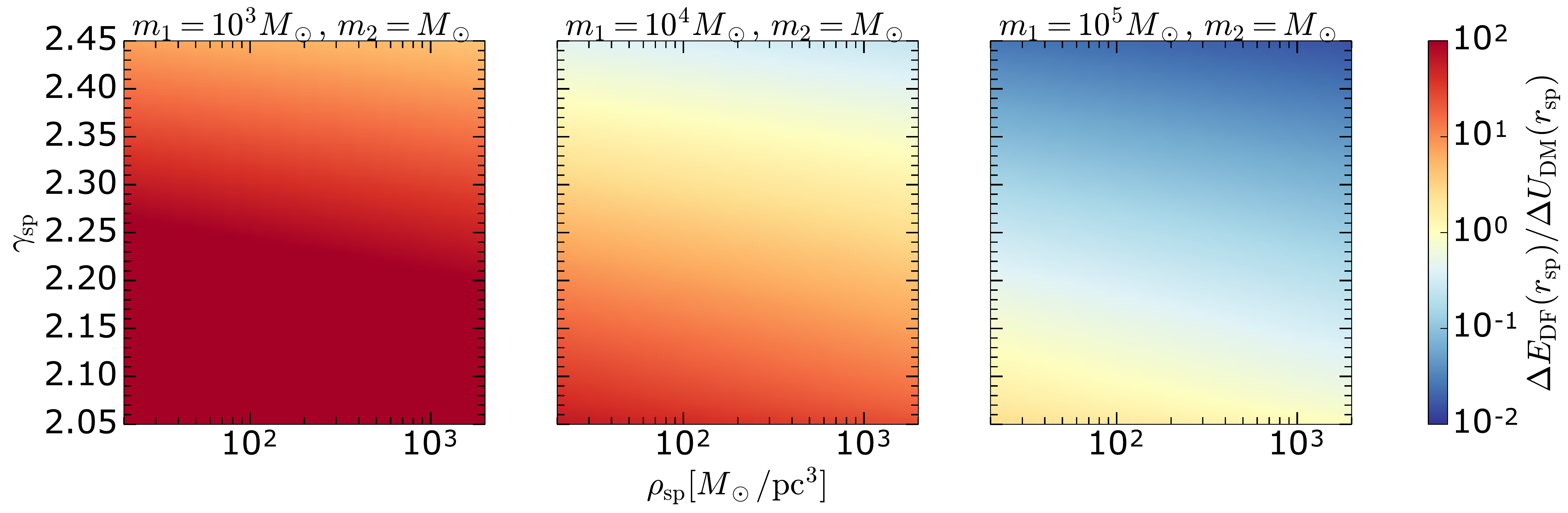}
    \caption{\textbf{Ratio of energy radiated through dynamical friction to binding energy of the DM spike for a range of DM spikes}.
    The three panels from left to right are the mass ratios
    $q=10^{-3}$, $10^{-4}$, and $10^{-5}$.
    The implications of this figure are discussed in more detail in Sec.~\ref{subsec:EnergyRatio}.}
    \label{fig:EDFbyUDMdensity}
\end{figure*}

Figure~\ref{fig:EDFbyUDMcurves} shows just one specific DM spike,
but there is nothing particularly special about the values
$\rho_\rmsp = 226 M_\odot/\mathrm{pc}^3$ and $\gamma_\rmsp = 7/3$
that were selected.
To illustrate how the results in Fig.~\ref{fig:EDFbyUDMcurves}
change for different values of $\rho_\rmsp$ and $\gamma_\rmsp$,
we show in Fig.~\ref{fig:EDFbyUDMdensity} the same ratio 
$E_\rmDF(r_\rmsp)/\Delta U_\rmDM(r_\rmsp)$ for a range of 
DM densities $\rho_\rmsp$ and power laws $\gamma_\rmsp$.
The three images correspond to the same three mass ratios 
shown in Fig.~\ref{fig:EDFbyUDMcurves}.
From left to right, they are $q=10^{-3}$, $10^{-4}$, and $10^{-5}$.
There are some common trends in all three panels: more dense 
(larger $\rho_\rmsp$) and steeper (larger $\gamma_\rmsp$) spikes
tend to have smaller ratios 
$\Delta E_\rmDF(r_\rmsp) / \Delta U_\rmDM(r_\rmsp)$ (i.e.,
satisfy energy balance better).
Even over this larger parameter space of DM spikes, the 
binary with a $10^{-3}$ mass ratio does not have a region where
$E_\rmDF(r_\rmsp)/\Delta U_\rmDM(r_\rmsp) < 1$.
The $10^{-5}$ mass-ratio binary has 
$E_\rmDF(r_\rmsp)/\Delta U_\rmDM(r_\rmsp) < 1$ for most spike
parameters, while the $10^{-4}$ mass ratio binary has the most
variation about $E_\rmDF(r_\rmsp)/\Delta U_\rmDM(r_\rmsp) \approx 1$.

Thus, in many (though not all) of the systems considered by \EdaEtAl, there is more energy dissipation through dynamical friction than binding energy in the DM distribution to account for this dissipation.
It will therefore be necessary to modify the distribution of DM in response to the energy input into the DM spike through dynamical friction.

Before implementing such a prescription, it would be of interest to know whether there is enough binding energy in the DM distribution to have a significant impact on the evolution of the binary.
We introduce a simple effective model in Appendix~\ref{sec:ShellModel}, in which dynamical friction is assumed to unbind all particles in the DM spike at a given radius. This model suggests that there is indeed sufficient binding energy to have an important effect.
Thus, we next turn to a more detailed description of how we implement this feedback on the DM distribution. 

\section{N-body simulations}
\label{sec:Nbody}

In order to build a semi-analytic prescription for feedback in the DM spike, we need to study in more detail the physics of dynamical friction in IMRI systems. In particular, as we will see in Sec.~\ref{sec:HaloFeedback}, we need to know the minimum and maximum impact parameter, $b_\mathrm{min}$ and $b_\mathrm{max}$
to include in our calculation of the dynamical friction effect. It is also useful to verify that the standard Chandrasekhar prescription for dynamical friction (which is derived for uniform density distributions) applies also in our setup.

For concreteness, we fix the minimum impact parameter to be $b_\mathrm{min} = 10 \,\mathrm{km}$, roughly the radius of a neutron star~\cite{Ozel:2016oaf}. In principle, $b_\mathrm{min}$ could be smaller (for example, if the orbiting compact object is a black hole rather than a neutron star). However, we do not need to worry about the precise value; these $\mathcal{O}(\mathrm{km})$ scales are much smaller than any other length scales in the problem and  can effectively be set to zero.

Instead, fixing the value of the maximum impact parameter $b_\mathrm{max}$ is crucial, as it determines which DM particles in the spike interact gravitationally with the orbiting compact object and therefore governs how energy is injected into the spike. 
Fixing $b_\mathrm{max}$ can also be seen as fixing the Coulomb logarithm $\log\Lambda$, because for $b_\mathrm{min} \rightarrow 0$, Eq.~\eqref{eq:coulombLog} becomes:

\begin{equation}
    \label{eq:Lambda1}
    \log \Lambda \approx \log \left( \frac{b_\mathrm{max}}{b_{90}}\right)\,.
\end{equation}

For the systems we consider here, a range of values have been previously assumed for the Coulomb logarithm. Reference~\cite{Hannuksela:2019vip} set $b_\mathrm{max}$ equal to the orbital radius, which would be appropriate for the motion of a compact object through a diffuse host such as a galaxy~\cite[p.664]{BinneyAndTremaine}. For a mass ratio of $q = 10^{-4}$ and an orbital radius of $20 G m_1/c^2$, this gives $\log \Lambda \sim 3$, the value used by \EdaEtAl~\cite{Eda:2013gg,Eda:2014kra}. For a compact object orbiting around a central point mass, we can combine Eqs.~\eqref{eq:Lambda1} and \eqref{eq:b90}, to show that $\log\Lambda = \log(1/q)$ under these assumptions.

However, the dynamics of DM particles at small radii will be dominated by the central IMBH, so it seems implausible that these particles can be deflected by the smaller orbiting compact object. A more plausible approach then is to fix $b_\mathrm{max}$ the distance at which perturbations from the small compact object can become relevant. The gravitational force from the central BH and from the compact object will be equal at a distance:
\begin{equation}
    b_\mathrm{max} \approx \sqrt{\frac{m_2}{m_1}}r_2\,,
\end{equation}
from the compact object. The corresponding Coulomb logarithm would then be:
\begin{equation}
    \log \Lambda = \log\sqrt{\frac{1}{q}} =\log\sqrt{\frac{m_1}{m_2}}\,.
\end{equation}

In order to determine the value of the maximum impact parameter, we perform a number of simulations using the publicly available \textsc{Gadget}-2 code \cite{Springel:2000yr,Springel:2005mi} as a pure $N$-body solver. For each simulation, we initialize a binary on a circular orbit with mass ratio $q = 10^{-3}$--$10^{-2}$, as well as a DM spike in dynamical equilibrium consisting of $N = 2^{15}$ particles. We evolve the system forward several hundred orbits and follow the evolution of the orbital separation. This allows us to calibrate the dynamical friction force and therefore determine $\log\Lambda$ and $b_\mathrm{max}$. In all simulations, we use as a benchmark a DM spike with $\rho_\mathrm{sp} = 226 \,M_\odot/\mathrm{pc}^3$ and a slope of $\gamma_\mathrm{sp} = 7/3$. Further details about the $N$-body simulations are given in Appendix~\ref{app:NbodyDetails}.

Figure~\ref{fig:separation} shows the change in orbital separation of the binary for a mass ratio $q = 10^{-2}$ and initial separation $r_2 = 3 \times 10^{-8}\,\mathrm{pc}$. Each curve shows the simulation result for a different random realization of the DM spike. These simulations cover approximately 3 days in physical time and take roughly the same length of time to simulate on 16 cores. Such simulations are therefore not suitable to follow the full evolution of the binary over many years, but do allow us to measure the size of the dynamical friction losses from the change in orbital energy:
\begin{equation}
\label{eq:DF_energyloss}
    \frac{\mathrm{d} E_{\mathrm{DF}}}{\mathrm{d} t} \approx \frac{G m_1 m_2}{2 (r_2)^2}\frac{\Delta r_2}{\Delta t}\,.
\end{equation}
For each binary configuration, we run at least 5 simulations, each for at least 100 orbits. The rate of dynamical friction energy loss in each simulation is estimated using Eq.~\eqref{eq:DF_energyloss}. This allows us to estimate the mean energy loss rate, as well as the error associated with different random realizations of the DM spike.

\begin{figure}[tb!]
    \centering
    \includegraphics[width=1.0\linewidth]{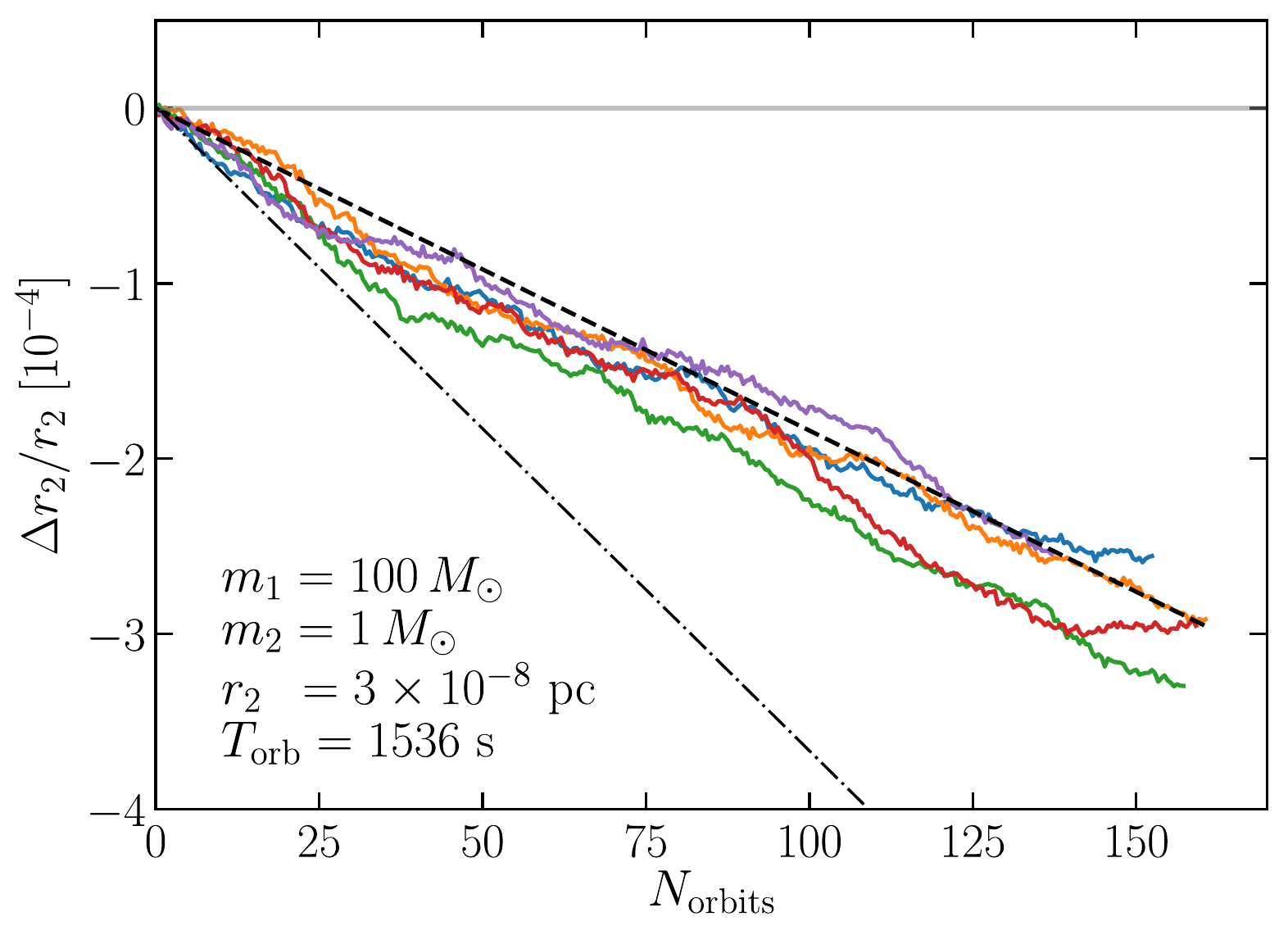}
    \caption{\textbf{Change in binary separation.} We show the results of 5 $N$-body simulations which are identical except for having different initial random realizations of the DM halo. The black dashed line shows the expected change in binary separation $r_2$, assuming dynamical friction losses as in Eq.~\eqref{eq:DFdissipation} and assuming $\Lambda = \sqrt{m_1/m_2}$, while the black dot-dashed line shows the expectation for $\Lambda = m_1/m_2$.}
    \label{fig:separation}
\end{figure}

\begin{figure}[t!]
    \centering
    \includegraphics[width=1.0\linewidth]{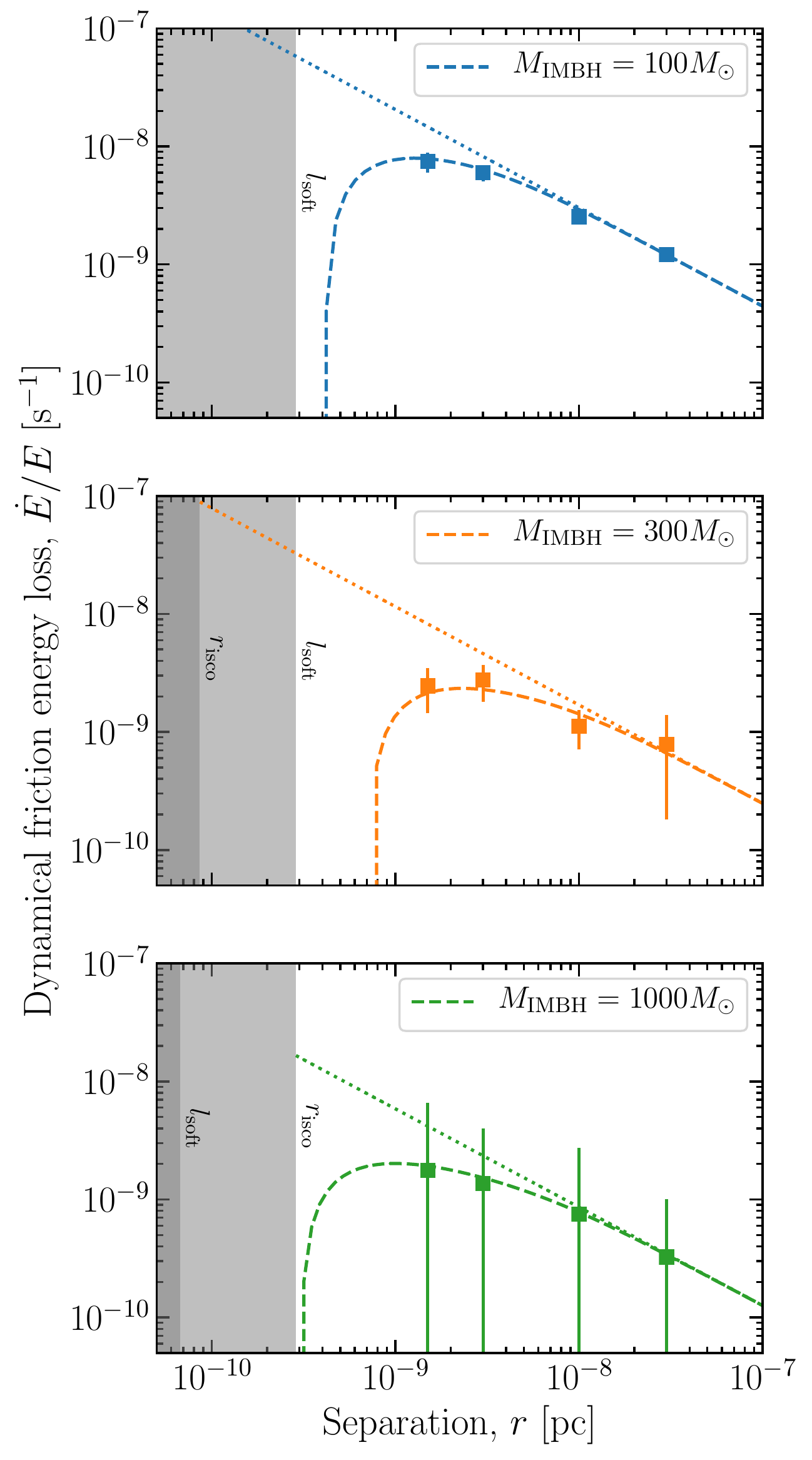}
    \caption{\textbf{Dynamical friction energy loss estimates from $N$-body simulations.} The orbiting compact object has a mass $m_2 = 1\,M_\odot$ and we show results for three masses of the central black hole: $100 \,M_\odot$ (\textbf{top}), $300 \, M_\odot$ (\textbf{middle}) and $1000 \,M_\odot$ (\textbf{bottom}). The diagonal dotted line shows the predicted energy loss from Eq.~\eqref{eq:DFdissipation}, assuming $\Lambda = \sqrt{m_1/m_2}$. The curved dashed line shows the energy loss accounting for the finite softening length. We also highlight in each panel the innermost stable circular orbit $r_\mathrm{isco}$ of the central IMBH, as well as the softening length $l_\mathrm{soft}$ of the simulations.}
    \label{fig:DF_panels}
\end{figure}

Figure~\ref{fig:DF_panels} shows the fractional energy-loss rate due to dynamical friction for binaries with central BH mass $m_1 = 100 \,M_\odot$ (top panel), $m_1 = 300 \,M_\odot$ (middle panel) and $m_1 = 1000 \,M_\odot$ (bottom panel). The dotted line in each panel shows the physical energy loss rate assuming $\log \Lambda = 1/\sqrt{q}$.  
For the systems we are studying, we can typically set $b_\mathrm{min} \rightarrow 0$, as discussed above. However, the simulations have a different minimum impact parameter due to their finite softening lengths $l_\mathrm{soft}$. The dashed lines show the energy loss rate which we expect to observe in the simulations, taking into account this finite softening lengths. The data points are well fit by $b_\mathrm{min} = \frac{1}{2}l_\mathrm{soft}$. As we move towards smaller separations, the maximum impact parameter shrinks, as the gravitational influence of the central BH increasingly dominates. At some point, the maximum impact parameter becomes comparable to the softening length of the simulations and the dynamical friction effect is no longer observable, shown as a sharp drop-off in the dashed curve.\footnote{Using the same logic, the maximum impact parameter is smaller for $m_1 = 1000\,M_\odot$ than for the less massive central black holes. In the case of $m_1 = 1000\,M_\odot$, we therefore use a slightly smaller softening length in order to preserve the dynamical friction effect down to smaller orbital separations.}

We see that in each panel of Fig.~\ref{fig:DF_panels}, the standard Chandrasekhar prescription for dynamical friction, for which we use $\Lambda = \sqrt{m_1/m_2}$, provides a good fit to the simulations. 
As we increase $m_1$, the uncertainties on the energy loss rate increase, as the central density of the spike grows. This in turn means that for a fixed number of DM pseudoparticles, the mass per pseudoparticle grows, giving a larger discretization noise in the simulations. Even so, the mean dynamical friction  effect estimated from the simulations matches Eq.~\eqref{eq:DFdissipation} well. This good match requires us to include the factor $\xi$, which accounts for the fraction of slow-moving DM particles and which was neglected in previous studies.

\begin{figure}[t!]
    \centering
    \includegraphics[width=1.0\linewidth]{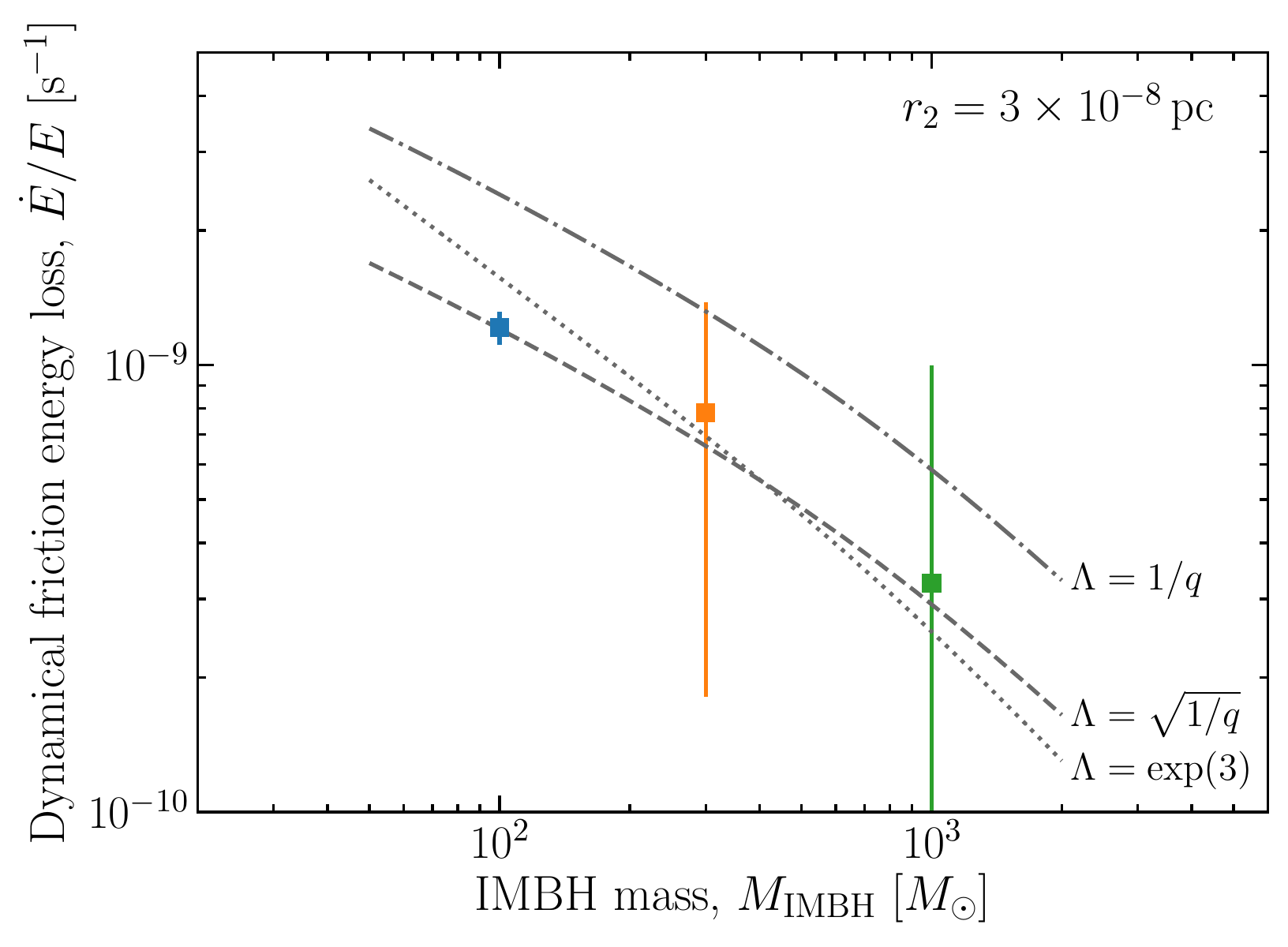}
    \caption{\textbf{Dynamical friction energy loss as a function of IMBH mass.} The data point for each of the three masses is the same as the right-most data point in the corresponding panel of Fig.~\ref{fig:DF_panels}. Lines correspond to the predicted rates of energy loss for three different values of the Coulomb factor $\Lambda$, where $q = m_2/m_1$.}
    \label{fig:DF_mass}
\end{figure}

In Fig.~\ref{fig:DF_mass}, we take the data points for $r_2 = 3 \times 10^{-8}\,\mathrm{pc}$ in each of the panels of Fig.~\ref{fig:DF_panels} and plot them together. We also plot the expected size of the dynamical friction loss for different values of the Coulomb term $\Lambda$. The best fit is provided by $\Lambda = \sqrt{1/q} = \sqrt{m_1/m_2}$, which was motivated by limiting the scattering to the gravitational sphere of influence of the orbiting compact object. We therefore use this value throughout the remainder of this paper, along with the corresponding value of $b_\mathrm{max}$:
\begin{equation}
\label{eq:bmax}
    b_\mathrm{max} = b_{90} \sqrt{\frac{m_1}{m_2}} = \sqrt{\frac{m_2}{m_1}} r_2\,.
\end{equation}

With these results, we can also verify the standard Chandrasekhar prescription for dynamical friction, which relies on the assumption of a uniform background distribution of scattering particles. Figure~\ref{fig:DF_panels} shows already that the dynamical friction correctly traces the DM density as a function of orbital radius, despite the fact that the DM distribution is not uniform. From Eq.~\eqref{eq:bmax},  $b_\mathrm{max}/r_2 = \sqrt{m_2/m_1}$, meaning that for a mass ratio of $q = 10^{-3}$, $b_\mathrm{max}$ is some 30 times smaller than the binary separation. The dynamical friction process therefore takes place only over a small region close to the orbiting compact object.
This implies that it is consistent to model the dynamical friction force for nonuniform systems using the local density, and it further implies that the dephasing effect could be used to accurately trace out the density profile as a function of radius.

\section{Halo feedback}
\label{sec:HaloFeedback}

We now describe a prescription to incorporate feedback in the DM spike, which we then use in Sec.~\ref{sec:Results} to follow the evolution of the binary self-consistently. This prescription is semi-analytic and allows us to track the phase space distribution of the DM spike as energy is injected by the inspiraling compact object. We begin by discussing the key assumptions behind our approach.

\begin{enumerate}[label=(\alph*)]
 
    \item We assume that the orbital elements evolve on a timescale that is long compared to the orbital period. This assumption is justified over most of the inspiral, as discussed below Eq.~\eqref{eq:Eorbit}, and allows us to consider the rate of energy being injected into the halo as constant over a small number of orbits. Note that we will not attempt to resolve changes to the distribution of DM on timescales shorter than a single orbit (because of the varying orbital phase of the compact object).
 
    \item We assume that the equilibration timescale for the DM halo is much shorter than the timescale for the secular evolution of the system. When DM particles in the halo receive a `kick' from the compact object, they move to a new orbit with a larger semi-major axis. It will thus take a few orbital periods before the distribution of these particles reflects the new equilibrium density profile. However, as discussed above, the evolution of the orbital elements is slow compared to these timescales. This allows us to compute the new equilibrium density profile of the DM `instantaneously' after energy is injected.

    \item We assume that the DM halo is spherically symmetric and isotropic, and remains so throughout the evolution of the system. This allows for a simpler treatment of the halo, as we need only keep track of the evolution of the energy of the DM particles and not their angular momentum. We discuss this assumption in more detail in Sec.~\ref{sec:Discussion}.

\end{enumerate}

With these assumptions, we can describe the DM in the spike at all times with an equilibrium phase space distribution function $f = m_\mathrm{DM} \mathrm{d} N/\mathrm{d}^3 \mathbf{r}\,\mathrm{d}^3\mathbf{v}$. If the distribution of DM is spherically symmetric and isotropic, then $f = f(\calE)$ and depends only on the relative energy per unit mass:
\begin{equation}
    \calE(r,v) = \Psi(r) - \frac{1}{2}v^2\,.
\end{equation}
Here, $\Psi(r) = \Phi_0 - \Phi(r)$ is the relative potential, with $\Phi(r)$ the standard gravitational potential and $\Phi_0$ a reference potential. Gravitationally bound particles then correspond to those with $\calE > 0$. The orbital separations we are interested in lie well within the sphere of influence of the central IMBH. We therefore neglect the gravitational potential due to the DM halo and write $\Psi(r) = G m_1/r$ (see, e.g., Appendix~II of Ref.~\cite{Edwards:2019tzf} for a numerical comparison). This also allows us to assume that the DM halo evolves in a fixed gravitational potential, rather than having to update the potential as the DM halo is perturbed.

For a given density profile $\rho(r)$, the distribution function $f(\calE)$ can be recovered using the Eddington inversion procedure~\cite[p.~290]{BinneyAndTremaine}. The initial equilibrium distribution function of the power-law spike is given by~\cite{Edwards:2019tzf}:
\begin{align}
    \begin{split}
       f_i(\mathcal{E})&=\frac{\gamma_\mathrm{sp}(\gamma_\mathrm{sp}-1)}{(2 \pi)^{3 / 2}} \rho_{\mathrm{sp}}\left(\frac{r_{\mathrm{sp}}}{G  m_1}\right)^{\gamma_\mathrm{sp}} \\ &\quad\quad \times\frac{\Gamma(\gamma_\mathrm{sp}-1)}{\Gamma\left(\gamma_\mathrm{sp} -\frac{1}{2}\right)} \mathcal{E}^{\gamma_\mathrm{sp}-3 / 2}\,,     
    \end{split}
\end{align}
where $\Gamma$ is the complete Gamma function. For a given distribution function, the density can be recovered as:
\begin{align}
\label{eq:rho_r}
    \begin{split}
        \rho(r) = 4 \pi \int_0^{v_\mathrm{max}(r)} v^2 f\left(\Psi(r) - \frac{1}{2}v^2\right) \,\mathrm{d}v\,,
    \end{split}
\end{align}
where $v_\mathrm{max}(r) = \sqrt{2\Psi(r)}$ is the escape velocity at radius $r$. Thus, if we can study the evolution of the distribution function $f$, then we can self-consistently evolve the DM halo along with the binary and reconstruct the density profile, which is required to calculate the dynamical friction force. A similar approach to the evolution of DM around BHs was applied in Refs.~\cite{Bertone:2005hw,Merritt:2006mt}.

The number of particles with energies $\mathcal{E} \rightarrow \mathcal{E} + \mathrm{d}\mathcal{E}$ is:
\begin{equation}
    N(\mathcal{E})\,\mathrm{d}\mathcal{E} = \frac{1}{m_\mathrm{DM}}g(\mathcal{E}) f(\mathcal{E}) \,\mathrm{d}\mathcal{E}\,.
\end{equation}
The density of states $g(\mathcal{E})$ denotes the volume of phase space per unit energy \cite[p. 292]{BinneyAndTremaine}. In the potential of the central BH, this can be calculated as:
\begin{align}
    \begin{split}
    \label{eq:DensityOfStates}
        g\left(\mathcal{E}\right)&=\int  \delta\left(\mathcal{E}-\mathcal{E}(r, v)\right) \mathrm{d}^{3} \mathbf{r} \,\mathrm{d}^{3} \mathbf{v}\\
        &=16 \pi^{2}  \int_{0}^{r_{\mathcal{E}}} \mathrm{d} r r^{2} \sqrt{2\left(\Psi(r)-\mathcal{E}\right)}\\
        &= \sqrt{2}\pi^3 G^3 m_1{}^3 \mathcal{E}^{-5/2}\,,
    \end{split}
\end{align}
where $r_\mathcal{E} = G m_1/\mathcal{E}$ is the maximum radius for a particle of energy $\mathcal{E}$.\footnote{We note that formally $g(\mathcal{E})f(\mathcal{E})$ diverges as $\mathcal{E} \rightarrow 0$ for $\gamma_\mathrm{sp} < 4$. However, we have so far only considered a DM spike which extends out to infinity. In practice, the DM spike will be smoothly truncated at large radii, modifying the distribution function as $\mathcal{E} \rightarrow 0$ and ensuring that the total number of DM particles remains finite.} 

Let us write $P_\mathcal{E}(\Delta\mathcal{E})$ as the probability (over a single orbit) that a particle with energy $\mathcal{E}$ scatters with the compact object and gains an energy $\Delta\mathcal{E}$. Then, the change in the number of particles at energy $\mathcal{E}$ over a single orbit can be written as:
\begin{align}
\begin{split}
\label{eq:DeltaN}
    \Delta N (\mathcal{E}) &= - N(\mathcal{E}) \int P_\mathcal{E}(\Delta\mathcal{E}) \,\mathrm{d}\Delta\mathcal{E}\\
    &+ \int N(\mathcal{E} - \Delta \mathcal{E}) P_{\mathcal{E} - \Delta\mathcal{E}}(\Delta\mathcal{E})  \,\mathrm{d}\Delta\mathcal{E} \,,
\end{split}
\end{align}
where the integration is over the range $\left[\Delta \mathcal{E}(b_\mathrm{max}), \Delta\mathcal{E}(b_\mathrm{min})\right]$.
The first term in Eq.~\eqref{eq:DeltaN} describes those particles initially at energy $\mathcal{E}$ which scatter to another energy, while the second term corresponds to those particles which scatter from energies $\mathcal{E}-\Delta \mathcal{E}$ to energy $\mathcal{E}$.

We will describe the evolution of the system in terms of the distribution function $f(\calE) = m_\mathrm{DM} N(\calE)/g(\calE)$. Assuming that the evolution of the system is much slower than the orbital frequency, we can write $\Delta f \approx T_\mathrm{orb} \,\partial f/\partial t$, with $T_\mathrm{orb} = 2\pi\sqrt{(r_2)^3/(G M)}$ the orbital period. Thus, we obtain:
\begin{align}
\label{eq:dfdt}
    \begin{split}
        &T_\mathrm{orb} \frac{\partial f(\calE, t)}{\partial t} = - p_\mathcal{E}f(\mathcal{E}, t)  \,+ \\
        &\int \left(\frac{\calE}{\calE - \Delta\calE}\right)^{5/2} f(\calE - \Delta\calE, t)  P_{\calE-\Delta\calE}( \Delta\calE)\,\mathrm{d}\Delta\calE\,,
    \end{split}
\end{align}
where $p_\mathcal{E} = \int P_{\mathcal{E}}(\Delta\mathcal{E}) \,\mathrm{d}\Delta\mathcal{E}$ is the total probability for a particle of energy $\mathcal{E}$ to scatter with the compact object during one orbit. 
We note that while we do not write $P_\mathcal{E}(\Delta \mathcal{E})$ with an explicit time-dependence, this probability depends implicitly on time through the orbital velocity and orbital radius $r_2(t)$.  
Using Eq.~\eqref{eq:dfdt}, we can evolve the distribution function over a number of orbits (assuming that the binary separation changes slowly compared to the orbital period). The density profile throughout the spike can then be derived using Eq.~\eqref{eq:rho_r}, which in turn is used to evaluate the rate of energy loss due to dynamical friction, given in Eq.~\eqref{eq:DFdissipation}.

The final step is then to evaluate $P_\calE( \Delta \calE)$. When a DM particle passes the compact object with impact parameter $b$, it is deflected and the velocity of the compact object parallel to its motion changes.\footnote{Note that we do not consider changes in the velocity \textit{perpendicular} to the motion of the compact object because on average these do not give rise to a change in energy.} The change in speed of the compact object is \cite[App.~L]{BinneyAndTremaine}:
\begin{equation}
\Delta v_{\parallel} = -2 v_0 \frac{m_\mathrm{DM}}{m_2} \left[1 + \frac{b^2}{b_{90}{}^2}\right]^{-1}\,,
\end{equation}
where $v_0$ is the relative speed of the encounter and $b_\mathrm{90}$ was defined in Eq.~\eqref{eq:b90}.
The change in energy of the compact object is then
\begin{equation}
\Delta E_\mathrm{CO} = \frac{1}{2}m_2\left[(v_0 + \Delta v_{\parallel})^2 - v_0^2\right] \approx m_2 v_0 \,\Delta v_{\parallel}\,,
\end{equation}
meaning that by energy conservation the change in relative energy per unit mass $\mathcal{E}$ of a single DM particle is:
\begin{equation}
\label{eq:deltaE}
\Delta \mathcal{E}(b) = -\frac{\Delta E_\mathrm{CO}}{m_\mathrm{DM}}  = -2 v_0^2 \left[1 + \frac{b^2}{b_{90}{}^2}\right]^{-1}\,.
\end{equation}
In principle, encounters between DM particles and the compact object occur with a range of relative speeds (owing to the velocity distribution of DM). Here for simplicity we fix the encounter speed to be equal to the orbital speed $v_0 = v_\mathrm{orb}$. We assume that only DM particles with speeds slower than $v_0 = v_\mathrm{orb}$ will scatter and gain energy from the orbiting compact object \cite{Chandrasekhar1943a}. For an isotropic velocity distribution, these assumptions give the correct total dynamical friction force on the compact object~\cite[Sec.~8.1]{BinneyAndTremaine}. Note that particles moving faster than $v_0$ will instead give rise to dynamical heating, increasing the energy of the compact object. However, this effect is suppressed by the ratio $m_\mathrm{DM}/m_2$ and can safely be neglected in this scenario \cite[p. 582]{BinneyAndTremaine}.

The scattering probability can now be evaluated as:
\begin{align}
    \begin{split}
    P_\mathcal{E}(\Delta\mathcal{E}) &= \frac{1}{g(\mathcal{E})} \iint\displaylimits_{r < r_\calE, \, v < v_0} \delta\left(\calE(r,v) - \calE\right)\\
    &\qquad \times\delta\left(\Delta\calE(b) - \Delta\calE\right) \,\mathrm{d}^3\mathbf{r}\,\mathrm{d}^3\mathbf{v}\,.
    \end{split}
\end{align}
Evaluating the integral over $\mathbf{v}$, as in Eq.~\eqref{eq:DensityOfStates}, and using Eq.~\eqref{eq:deltaE} to change the argument of the second $\delta$-function, we obtain:
\begin{align}
    \begin{split}
    \label{eq:Ps_2}
        P_\mathcal{E}(\Delta\mathcal{E}) &= \frac{\pi b_{90}^2}{g(\calE) v_0^2} \int_{r_\mathrm{cut}}^{r_\calE} \frac{1}{b} \left[ 1 + \frac{b^2}{b_{90}{}^2}\right]^2 \\
        &\qquad \times\delta\left(b - b_\star(\Delta\calE)\right) \sqrt{2\left(\Psi(r)-\mathcal{E}\right)} \,\mathrm{d}^3\mathbf{r}\,.
    \end{split}
\end{align}
Here, we have defined $b_\star = b_{90} \sqrt{2 v_0^2/|\Delta\calE| - 1}$ and the lower limit $r_\mathrm{cut} = G m_1/(\mathcal{E} + \frac{1}{2}v_0^2)$ ensures that only particles with $v < v_0$ can scatter with the orbiting compact object.

\begin{figure}[tb]
    \begin{center}
    \includegraphics[width=0.99\columnwidth]{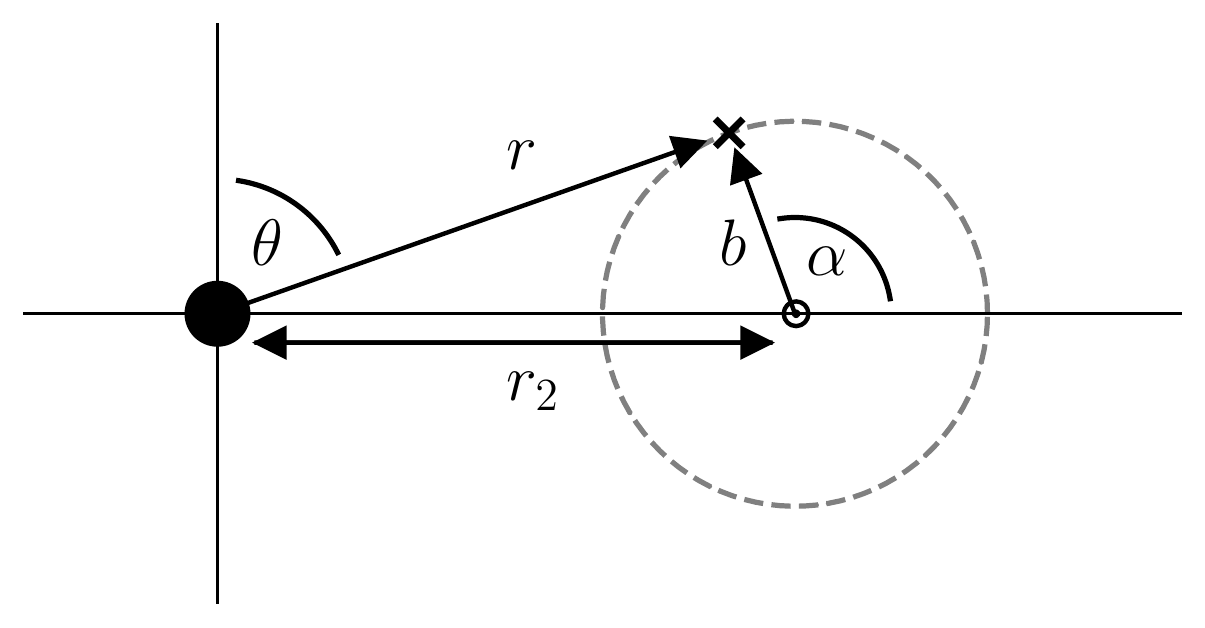}
    \end{center}
    \caption{\textbf{Geometry of DM scattering around the compact object.} The compact object position is denoted $\odot$, at a radius $r_2$ from the central IMBH. The motion of the compact object is into (or out of) the page. See Eqs.~\eqref{eq:Ps_2}-\eqref{eq:r} for more details.}
    \label{fig:Geometry}
\end{figure}

Equation~\eqref{eq:Ps_2} now involves an integral over the entire DM spike, with a contribution only from positions with impact parameters $b = b_\star(\Delta\calE)$. This corresponds to an integral over the torus with minor radius $b_\star(\Delta \calE)$ and major radius $r_2$, the orbital radius of the compact object. For $b \ll r_2$, we can perform the azimuthal integral over the orbit:
\begin{equation}
    \int r^2 \,\mathrm{d}r \,\mathrm{d}\cos\theta \,\mathrm{d}\phi \rightarrow 2\pi r_2 \int \sin\theta \,r \,\mathrm{d}r\, \mathrm{d}\theta\,,
\end{equation}
where $(r, \theta, \phi)$ are the standard spherical polar coordinates. Finally, we change variables from $(r,\theta)$ to $(b, \alpha)$, where the angle $\alpha \in [0, 2\pi]$ is defined as in Fig.~\ref{fig:Geometry}. With this, we have:
\begin{equation}
    \int  \sin\theta \,r \,\mathrm{d}r\, \mathrm{d}\theta \rightarrow 2\int_{0}^{\pi} \int_{0}^\infty  \sin\left(\theta[b, \alpha]\right)\, b\,\mathrm{d}b\,\mathrm{d}\alpha\,.
\end{equation}
Substituting in Eq.~\eqref{eq:Ps_2} and performing the integral over $b$, we finally obtain:
\begin{align}
    \begin{split}
    \label{eq:Ps_3}
        P_\mathcal{E}(\Delta\mathcal{E}) &= \frac{4\pi^2  r_2}{g(\calE)}\frac{b_{90}{}^2}{v_0^2} \left[ 1 + \frac{b_\star^2}{b_{90}{}^2}\right]^2 \times \\ 
        & \int
         \sqrt{2\left(\Psi(r[b_\star, \alpha])-\mathcal{E} \right)} \sin\left(\theta[b_\star, \alpha]\right)\,\mathrm{d}\alpha\,.
    \end{split}
\end{align}
Note that here the value of $b_\mathrm{max} = r_2\sqrt{m_2/m_1}$ discussed in Sec.~\ref{sec:Nbody} sets the minimum value of $\Delta \mathcal{E}$, through Eq.~\eqref{eq:deltaE}. The radial coordinate $r$ is now expressed as:
\begin{equation}
\label{eq:r}
    r[b_\star, \alpha] = \left[ r_2^2 + b_\star^2 + 2r_2 b_\star \cos\alpha   \right]^{1/2}\,,
\end{equation}
and we integrate over all values of $\alpha \in [0, \pi]$ such that $r[b_\star, \alpha] \in [r_\mathrm{cut}, r_\calE]$. We work to first order in $b_\star/r_2$, in which case Eq.~\eqref{eq:Ps_3} can be written in terms of elliptic integrals; more details are provided in Appendix~\ref{app:Scattering}.  Code for computing the properties and time evolution of the DM spike is publicly available online at \href{https://github.com/bradkav/HaloFeedback}{https://github.com/bradkav/HaloFeedback} \cite{HaloFeedback}.

\subsection{Testing the halo feedback}
\label{sec:testing}

Before tackling the complete IMRI system including a dynamic DM spike, we first test the formalism by following the evolution of the DM distribution in a simpler scenario. We consider a mass $m_2 = 1.4 \,M_\odot$ orbiting a central BH $m_1 = 1400 \,M_\odot$ at a distance $r_2 = 10^{-8}\,\mathrm{pc}$. This configuration is a typical snapshot of an IMRI signal which would be observable by LISA, except that we will keep the orbital separation \textit{fixed}. That is, we will look only at how the DM spike evolves in response to energy injection, without allowing the orbit of the compact object to change.

\begin{figure}[t!]
    \begin{center}
    \includegraphics[width=0.99\columnwidth]{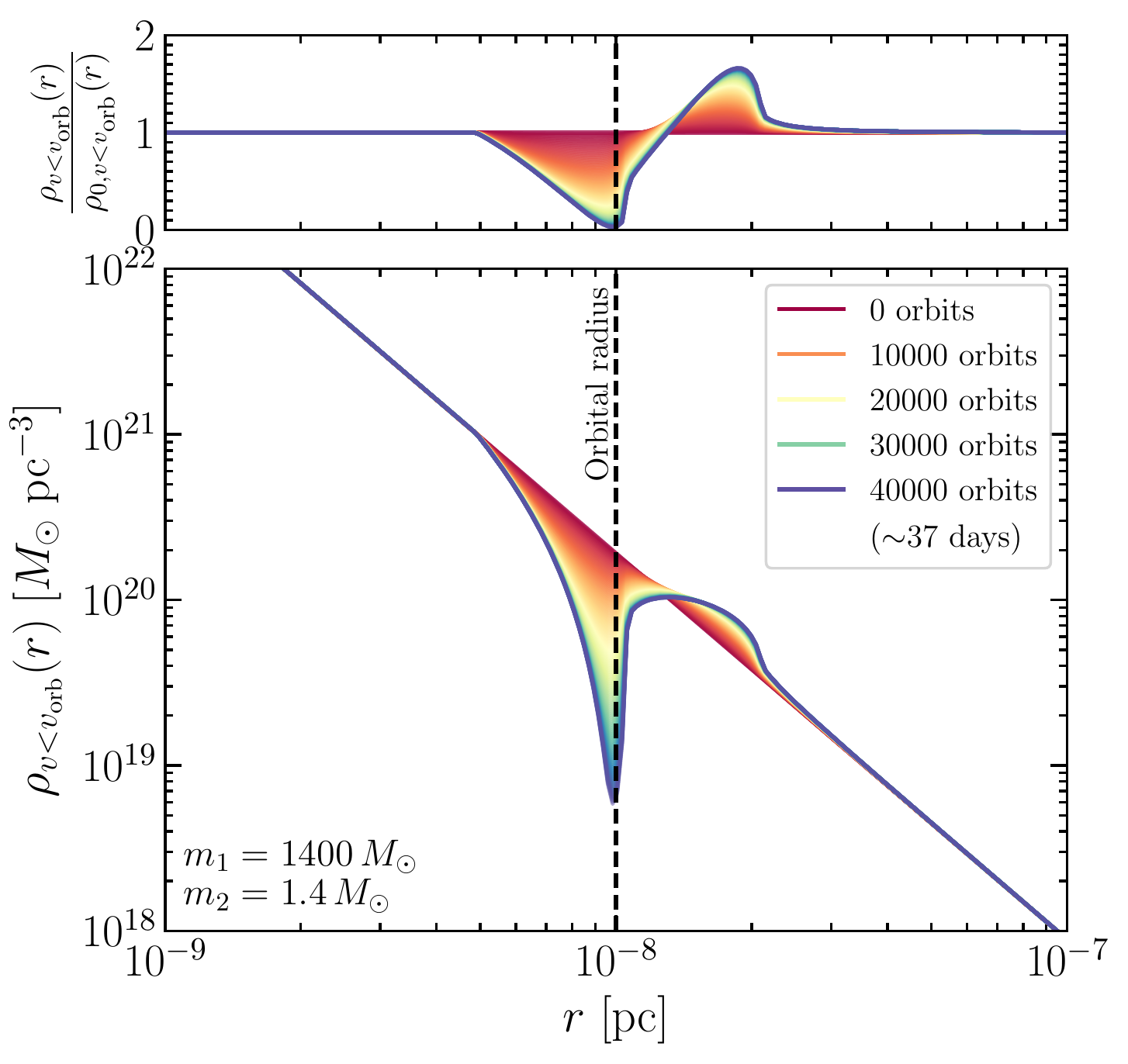}
    \end{center}
    \caption{\textbf{Evolution of the DM spike density profile due to feedback from the orbiting object.} We consider a compact object $m_2 = 1.4 \,M_\odot$ orbiting at a \textit{fixed} radius $r = 10^{-8}\,\mathrm{pc}$ from the IMBH with $m_1 = 1400\,M_\odot$. Note that we plot $\rho_\mathrm{DM}(r)$ multiplied by the fraction of DM at radius $r$ moving more slowly than the local orbital speed $v_\mathrm{orb}(r)$. The upper panel shows the evolution of the density profile normalised to the density profile $\rho_0$ at the start of the simulation.}
    \label{fig:Feedback}
\end{figure}

Figure~\ref{fig:Feedback} shows the result of this ``test'' simulation, run over 40000 orbits. We plot the density profile of the spike, including only those particles moving more slowly than the local orbital speed $v < v_\mathrm{orb}(r)$ (i.e.~only those particles which would produce a net dynamical friction effect on the orbiting compact object). DM particles are gradually depleted from close to the compact object through scattering; at the end of the simulation, the density at the orbital radius has dropped to 3\% of the initial density. We note that particles with some energy $\mathcal{E}$ naturally populate radii between $r = 0$ and $r = r_{\mathcal{E}} = G m_1/\mathcal{E}$. This means that particles scattering at a radius $r_2$ will also deplete particles at smaller radii, as observed in Fig.~\ref{fig:Feedback}. These scattered particles gain energy and their average radius increases, leading to a bump in the density profile at $r > r_2$.

By comparing the change in energy of the DM spike and the work which would be done on the compact object by dynamical friction, we have confirmed that energy is conserved at the level of 0.1\%, with approximately 6\% of the total energy carried away by particles which become completely unbound from the spike. Unlike in the case of a static DM spike, this feedback formalism allows us to follow the system self-consistently, ensuring that energy is conserved throughout its evolution.

\section{Evolution of the binary with halo feedback}
\label{sec:Results}

In this section, we incorporate the halo feedback prescription in Sec.~\ref{sec:HaloFeedback} into our evolution of the binary system.
We first discuss the evolution equations and our numerical methods for solving these equations; we then discuss the results of our numerical simulations.

\subsection{Evolution equations and numerical methods}
\label{subsec:EqsAndNumerics}

In Sec.~\ref{sec:EnergyBalance}, we could determine the dissipative dynamics of the binary from solving a single, ordinary differential equation for the orbital separation of the binary, Eq.~\eqref{eq:r_dot}, in a static distribution of dark matter, Eq.~\eqref{eq:rhoDM}.
In this section, we instead simultaneously evolve the orbital separation of the binary in a spherically symmetric distribution of dark matter, which itself evolves in response to the inspiral of the small compact object from dynamical friction.
Thus, the evolution equation for $\dot r_2$ has a similar form to that in Eq.~\eqref{eq:r_dot}, but we replace $\rho_\rmDM(r_2)$ with the time-dependent DM distribution evaluated at $r_2$, which we denote $\rho_\rmDM(r_2,t)$. Similarly, the fraction of DM particles slower that the circular speed at $r_2$ is written $\xi(r_2, t)$.
The expression, in full, is
\begin{equation} \label{eq:r_dot_feed}
\begin{split}
\dot{r}_2 = & - \frac{64\, G^3\, M \, m_1\, m_2}{5\, c^5\, (r_2)^3} \\
& - \frac{8 \pi\, G^{1/2}\, m_2 \, \log\Lambda  r_2^{5/2} \, \rho_\rmDM(r_2,t) \,\xi(r_2, t)}{\sqrt{M} m_1 }  \, .
\end{split}
\end{equation}
Because the evolution of the DM spike at all radii $r$, $\rho_\rmDM(r,t)$, depends upon $r_2$ we must simultaneously evolve Eq.~\eqref{eq:r_dot_feed} with the prescription in Sec.~\ref{sec:HaloFeedback} for evolving the dark-matter distribution.

Thus, the evolution equations that we must solve take the form of a coupled system of an ordinary differential equation and an integro-partial differential equation.
Schematically, the system has the form
\begin{subequations} \label{eq:ODE_PDEsystem}
\begin{align}
    \frac{\mathrm{d} r_2(t)}{\mathrm{d}t} = & F_1\left[r_2, \left[\int \mathrm{d}^3v f(\mathcal E, t; r_2)\right]_{r=r_2}\right] \, ,\\
    \frac{\partial f(\mathcal E, t; r_2)}{\partial t} = &
    F_2\left[f(\mathcal E, t; r_2), \int 
    \mathrm{d}\Delta\mathcal E
    f(\mathcal E - \Delta \mathcal E, t; r_2) \right] \, ,
\end{align}
\end{subequations}
where the explicit forms of the functionals $F_1$ and $F_2$ can be obtained from Eqs.~\eqref{eq:dfdt} and~\eqref{eq:r_dot_feed} [as well as the relationship given in Eq.~\eqref{eq:rho_r}].
Here we also added an explicit dependence of $f(\mathcal E, t)$ on $r_2$ using the notation $f(\mathcal E, t; r_2)$, so as to emphasize that the ordinary and partial differential equations are coupled.
When discretizing the system in Eq.~\eqref{eq:ODE_PDEsystem} to solve it numerically, we first use Simpson's rule to evaluate the integrals, and then we use the method of lines (discretizing the partial differential equation on a grid of $\mathcal E$ values and solving the resulting system of ordinary differential equations on these grid points) and a second-order-accurate Runge-Kutta method to numerically solve the coupled ordinary and integro-partial differential equations.
Because there are only integrals rather than derivatives appearing on the right-hand side of the partial differential equation, we did not find that there was a Courant-Friedrichs-Lewy condition~\cite{1928MatAn.100...32C} that limited the size of our timestep (unlike for explicit numerical schemes for solving the advection equation, for example).

There are also two somewhat subtle issues that arise when evolving the binary with the halo feedback, which are related to (i) initial conditions and (ii) the size of the time steps used to evolve the system.
We discuss each of these issues now in more detail.

(i) Regarding initial conditions, for simplicity, one might like to be able to use the static DM distribution, Eq.~\eqref{eq:rhoDM}, as the initial condition  for evolving the binary with halo feedback.
However, unless the small compact object suddenly materialized in its orbit, this will generally not be a realistic initial condition.
Rather, one would expect that the small compact object was either captured, or it formed at a larger radius, and altered the dark-matter distribution via feedback on the halo until it reaches an orbital separation from where it could be detectable by LISA.
This could make simulating the binary challenging, because the exact initial conditions could depend upon the history of how the binary formed.

However, as we saw in Sec.~\ref{sec:Nbody}, the particles that contribute to this gravitational drag force lie within some small range of impact parameters from the compact object. 
We anticipate then that outside of some distance from the small compact object, the distribution of dark matter is not strongly affected, and the static distribution of dark matter, Eq.~\eqref{eq:rhoDM}, remains a good approximation for the density within this region.
If we are interested in evolving the binary using more realistic initial conditions for an initial separation $r_\rmi$, then we would need to start evolving the system from a larger separation $r_\rmi + \Delta r_\rmi$, where we have defined $\Delta r_\rmi$ to be the distance outside of which the distribution of dark matter is not significantly affected by the gravitational scatterings that produce dynamical friction.
We will take this approach described here to set what we believe to be reasonable initial conditions for the evolution of the binary and the dark-matter spike; in practice, we set $\Delta r_\mathrm{i} = 2 r_\mathrm{i}$.

(ii) Regarding the size of time steps, we note that the method of Sec.~\ref{sec:HaloFeedback} for evolving the dark-matter halo is only valid over timescales of (at least) a few orbital periods.
Thus, we will be limited in the size of the time steps that we can take to be this size or greater.
While this will not be problematic when the system is adiabatically evolving between circular orbits, our errors could be large when the binary is more relativistic, and the orbital radius changes more rapidly.
Ultimately, we do not view this as a large problem, because the Newtonian approximation that we adopt throughout this paper runs into other inaccuracies when the system is sufficiently relativistic that we would like to be taking a smaller time step. 
Also, the dynamical friction effect is of a negative post-Newtonian order for quasicircular binaries, meaning that it is largest when the binary is less relativistic.
We discuss these issues in more detail in Sec.~\ref{sec:Discussion}. Nevertheless, because we can only take timesteps that are an integral number of the orbital periods, we will not be able to resolve the orbital phase (or changes in phase) to less than a few integral multiples of $2\pi$ (i.e., less than a few orbits).

We check the accuracy of our numerical methods through two types of tests.
First, to determine whether taking timesteps that are an integral number of the orbital period has an affect on our solving Eq.~\eqref{eq:r_dot} for a static DM distribution, we compare our numerical solution for the number of GW cycles as a function of the GW frequency to the analytical expression in Eq.~\eqref{eq:NcyclesStaticDM}.
We find that we can resolve the number of GW cycles to 10s of cycles.
Second, we ran numerical simulations of the dynamic DM spike at several different numerical resolutions (we considered a sequence of timesteps that were a different number of orbital periods) for the binary with $q=10^{-3}$ and the DM spike with the initial DM spike given by $\rho_\rmsp=\,226 M_\odot/\mathrm{pc}^3$ and
$\gamma_\rmsp = 7/3$.
We found by comparing the two highest resolutions that the accuracy of our simulations was more of order of 100 GW cycles.

\subsection{Results of numerical simulations}

First, we will qualitatively describe the behavior of the binary with a dynamical DM spike.
As we saw in Sec.~\ref{sec:testing}, feedback on the DM halo leads to a depletion of the DM density at the orbital radius. This in turn reduces the size of the dynamical friction force and thus slows the inspiral. There is therefore competition between how quickly the compact object depletes DM and how rapidly dynamical friction causes it to lose energy. If the inspiral is sufficiently fast, the compact object moves to an orbit at smaller radius before much of the DM is depleted and the overall effect of feedback will be relatively small. Instead, if the inspiral is slow, most of the DM will be depleted at the current orbital radius and the binary will effectively stall. At this stage GW energy losses become more significant, and the binary must move slowly to a smaller radius before dynamical friction can dominate again. In this case, the behavior of the system is significantly altered by feedback. Animations showing the co-evolution of the binary and DM profile are available online at \url{https://doi.org/10.6084/m9.figshare.11663676} \cite{Animations}. 

\begin{figure}[t!]
    \centering
    \includegraphics[width=\linewidth]{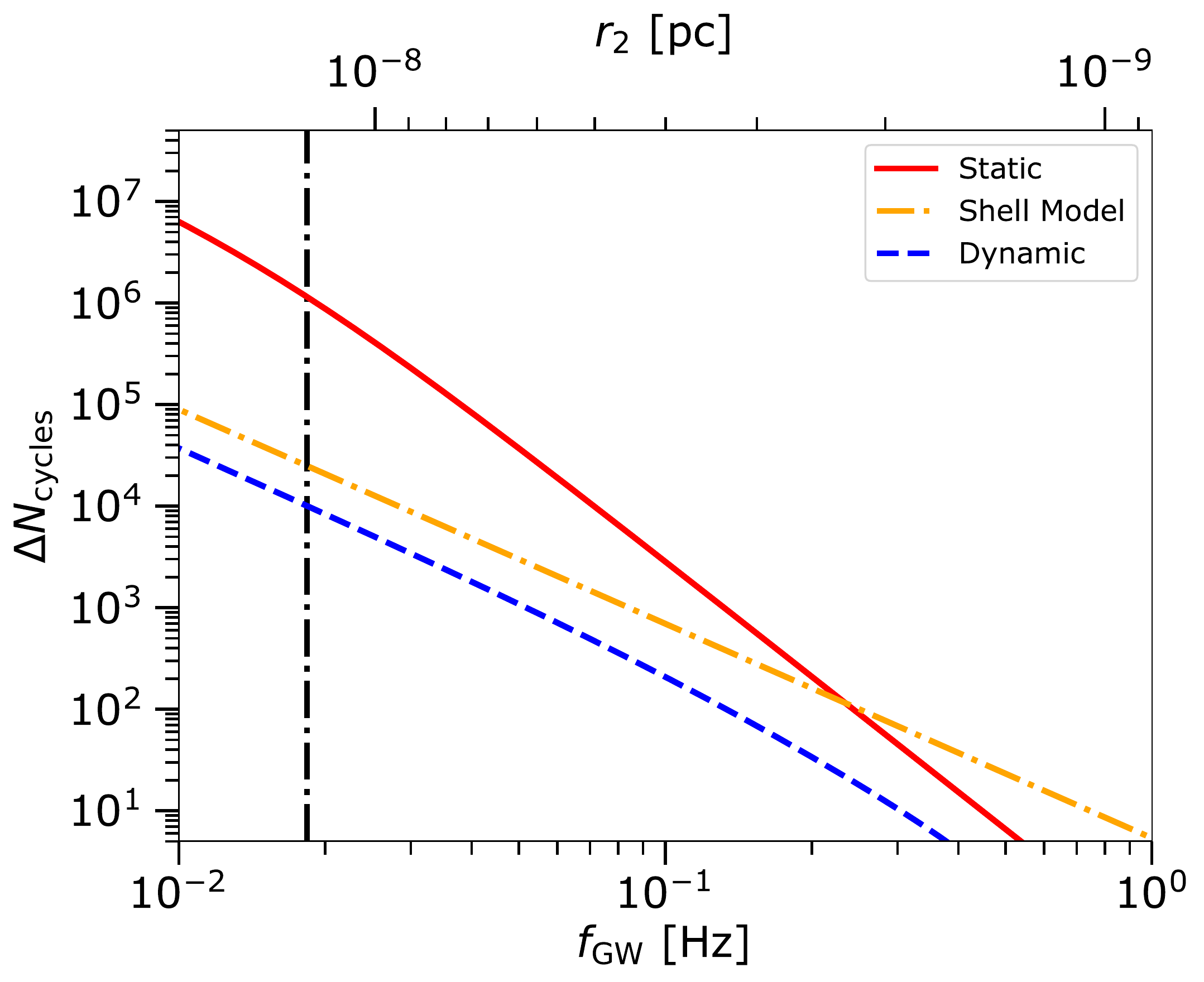}
    \caption{\textbf{Change in the number of GW cycles with respect to the vacuum inspiral.}
    For a mass ratio $q = 10^{-3}$, $m_1=1.4 \times 10^3\,M_\odot$, $\rho_\rmsp=226 \,M_\odot/\mathrm{pc}^3$ and $\gamma_\rmsp=7/3$, we show the change in the number of cycles (compared to the case without DM) starting from GW frequency $f_{\mathrm{GW}, i}$ up to the merger.
    The three curves show the change in cycles for a static DM distribution (solid red), the dynamic DM distribution (blue dashed) and a shell model (orange dotted-dashed) described in Appendix~\ref{sec:ShellModel}.
    The vertical dotted-dashed black line shows the GW frequency such that the system without DM will inspiral and merge within five years.}
    \label{fig:deltaN-cycles}
\end{figure}

To quantify the size of the dephasing effect, we estimate the difference between the number of gravitational wave cycles $N_\mathrm{cycles}$ during the inspiral in vacuum and in presence of the DM mini-spike, for both the dynamic and static cases. We define the number of GW cycles by integrating the GW frequency between two times,
\begin{equation} \label{eq:NcyclesTime}
N_\mathrm{cycles}(t_\mathrm{f}, t_\mathrm{i}) = 
\int^{t_\mathrm{f}}_{t_\mathrm{i}} f_\rmGW(t) \mathrm{d}t
\, .
\end{equation}
In the quadrupole approximation, the GW frequency $f_\rmGW$ is twice the orbital frequency $\Omega_\mathrm{orb}(t)/(2\pi)$. The GW frequency grows monotonically with time during the inspiral and we can therefore also express the number of cycles in terms of the initial and final GW frequencies:  $N_\mathrm{cycles}(f_{\mathrm{GW},_\mathrm{f}},f_{\mathrm{GW},_\mathrm{i}})$.

In Fig.~\ref{fig:deltaN-cycles}, we show the difference in the number of GW cycles with and without DM,
\begin{equation} \label{eq:NcyclesFreq}
    \Delta N_\mathrm{cycles} = N_\mathrm{cycles}^\mathrm{vac}(f_\mathrm{GW,f}, f_\mathrm{GW,i}) -N_\mathrm{cycles}^\mathrm{DM}(f_\mathrm{GW,f}, f_\mathrm{GW,i})\,,
\end{equation}
for a binary with masses $m_1 = 1400 \,M_\odot$ and $m_2=1.4 \,M_\odot$, and a fiducial spike with $\rho_\mathrm{sp} = 226 \,M_\odot/\mathrm{pc}^3$ and slope $\gamma_\mathrm{sp}= 7/3$.\footnote{Our choice of the mass $m_2$ is motivated by the Chandrasekhar limit~\cite{1931ApJ....74...81C,1931MNRAS..91..456C,1935MNRAS..95..207C}, though our results do not depend on the nature of either compact object (e.g.\ neutron star or black hole).} We fix the final frequency as the GW frequency at the ISCO $f_{\mathrm{GW,f}} \approx 3.1 \,\mathrm{Hz}$ and show $\Delta N_\mathrm{cycles}$ as a function of  $f_{\mathrm{GW,i}}$.\footnote{Note that this implies that the time it takes for the system to inspiral between the initial and final GW frequencies will differ for the system with and without DM.}
The solid red line shows results for a static DM spike (as described in Sec.~\ref{sec:EnergyBalance}). 
The dot-dashed orange line instead shows results for the model in Appendix~\ref{sec:ShellModel}, which we refer to as the ``shell model.''
In this model, the rate of dynamical friction energy loss is set equal to binding energy in the DM spike at any given radius. 
This toy model respects energy conservation and corresponds to the case where dynamical friction is maximally efficient, in the sense that all of the DM halo's binding energy is converted into work by dynamical friction. It is clear from Fig.~\ref{fig:deltaN-cycles} that the maximum allowed size of the dephasing effect, obtained in this toy model, can be as much as two orders of magnitude smaller than that estimated in the static case. The dashed blue curve shows our results for the dynamic DM spike, obtained using the prescription described in Sec.~\ref{sec:HaloFeedback}. 

At the lower range of the frequencies depicted in Fig.~\ref{fig:deltaN-cycles}, $\Delta N_\mathrm{cycles}$ for the dynamic spike is a factor of a few smaller than $\Delta N_\mathrm{cycles}$ for the shell model; however, as functions of frequency, both cases roughly follow the same power law.
This suggests that at lower frequencies (before GW energy loses become more efficient than loses from dynamical friction) the effects of dynamical friction on the orbital dynamics of the binary are similar to unbinding a fraction of a shell of DM particles at the orbital radius.
At the higher range of frequencies shown, $\Delta N_\mathrm{cycles}$ for the dynamic case follows a power law closer to that for the static case, but again it is a factor of a few smaller than the result for the static DM spike.
The following argument can explain this result: for the higher frequencies shown, GWs are more efficient in causing the binary to inspiral; thus, dynamical friction is not able to significantly change the DM spike and the dynamics of the system can be approximated well by having a static DM spike.
The magnitude of the dephasing is smaller in the dynamic case than in the static one, because the DM density is somewhat depleted by the effect of dynamical friction from earlier in the inspiral (\textit{cf.} the discussion of initial conditions in Sec.~\ref{subsec:EqsAndNumerics}).

\renewcommand{\arraystretch}{1.8}
\begin{table}[t]
\caption{ \textbf{Change in the number of cycles $\Delta N_\mathrm{cycles}$ during the inspiral.} Change in the total number of GW cycles  due to dynamical friction, starting 5 years from the merger. We compare results for a static DM halo and a dynamic DM halo incorporating feedback. In the top, middle and bottom tables, we show results for mass ratios of $q = 10^{-3}$, $10^{-4}$ and $10^{-5}$ respectively. We also indicate the number of cycles expected in vacuum (in the absence of DM). We fix $m_2 = 1.4\,M_\odot$ in all three cases. Note that $7/ 3 = 2.333\ldots \equiv 2.\overline{3}$.}

\begin{center}
  \begin{tabularx}{0.9\columnwidth}{lXXXX}
  \multicolumn{5}{c}{$m_1 = 1.4 \times 10^3\,M_\odot$, $N_\mathrm{cycles} = 4.63 \times 10^6$ in vacuum}
  \\ \hline\hline
                & $\gamma_\mathrm{sp} = 1.5$ & $\gamma_\mathrm{sp} = 2.2$ & $\gamma_\mathrm{sp} = 2.3$ & $\gamma_\mathrm{sp} = 2.\overline{3}$ \\\hline
        Static &     $\quad< 1$           &  $1.8 \times 10^4$    &  $1.1 \times 10^5$    & $2.1\times 10^5$ \\
       Dynamic &     $\quad< 1$           & $ 2.4 \times 10^2 $     &  $1.6 \times 10^3$    & $3.1 \times 10^3$ \\
       \hline\hline
  \end{tabularx}  

\vspace{\baselineskip}
  
    \begin{tabularx}{0.9\columnwidth}{lXXXX}
  \multicolumn{5}{c}{$m_1 = 1.4 \times 10^4\,M_\odot$, $N_\mathrm{cycles} = 2.60 \times 10^6$ in vacuum}
  \\ \hline\hline
        & $\gamma_\mathrm{sp} = 1.5$ & $\gamma_\mathrm{sp} = 2.2$ & $\gamma_\mathrm{sp} = 2.3$ & $\gamma_\mathrm{sp} = 2.\overline{3}$ \\\hline
       Static  &  $\quad< 1$ & $1.0 \times 10^3$ & $6.3 \times 10^3$& $1.2 \times 10^4$\\
        Dynamic & $\quad< 1$ & $5.0 \times 10^2$ & $3.1 \times 10^3$ & $5.8 \times 10^3$\\
        \hline\hline
  \end{tabularx}  

 \vspace{\baselineskip}
  
    \begin{tabularx}{0.9\columnwidth}{lXXXX}
  \multicolumn{5}{c}{$m_1 = 1.4 \times 10^5\,M_\odot$, $N_\mathrm{cycles} = 1.39 \times 10^6$ in vacuum}
  \\ \hline\hline
        & $\gamma_\mathrm{sp} = 1.5$ & $\gamma_\mathrm{sp} = 2.2$ & $\gamma_\mathrm{sp} = 2.3$ & $\gamma_\mathrm{sp} = 2.\overline{3}$ \\\hline
       Static  &  $\quad< 1$ & $5.5 \times 10^1$ & $3.3 \times 10^2$& $6.0 \times 10^2$\\
        Dynamic & $\quad< 1$ & $5.3 \times 10^1$ & $3.2 \times 10^2$ & $5.9 \times 10^2$\\
        \hline\hline
  \end{tabularx}  
  
\end{center}

\label{tab:Ncycles}
\end{table}

In Table~\ref{tab:Ncycles}, we list numerical values of $\Delta N_\mathrm{cycles}$ for different configurations of the IMRI system and DM spike. Having in mind a 5 year observation with LISA, we measure $\Delta N_\mathrm{cycles}$ starting from a separation (or, equivalently, an initial frequency) such that the time-to-merger is 5 years in the both the vacuum and DM cases. Note that this means that the systems with DM will start at a larger separation (or lower initial frequency) than the vacuum case, in order to give a merger in the same time.\footnote{For reference, for a $1.4 \times 10^3 \,M_\odot$ ($1.4 \times10^4 \,M_\odot$) IMBH, the initial separation giving a five year inspiral in the vacuum case is  $r_2 = 1.24 \times 10^{-8} \,\mathrm{pc}$ ($r_2 = 3.92 \times 10^{-8} \,\mathrm{pc}$).} Note that this convention for specifying $\Delta N_\mathrm{cycles}$ differs somewhat from the definition used in \EdaEtAl (their convention is equivalent to that used in Fig.~\ref{fig:deltaN-cycles}); however, because LISA will operate for a fixed amount of time, and because sources like IMRIs typically will not merge on a timescale shorter than that of LISA's operation, we opt to compare the number of cycles over a fixed time rather than from a fixed initial frequency.
These different conventions do change the difference in the number of cycles, so, for example, the results in Fig.~\ref{fig:deltaN-cycles} and the numbers in Table~\ref{tab:Ncycles} cannot be directly compared, even for the same binary and DM spike.

For a central IMBH with $m_1 = 1.4 \times 10^3 \,M_\odot$, assuming a static DM spike with slope $\gamma_\mathrm{sp} = 7/3$, the dephasing effect would reduce the number of GW cycles from the value in vacuum by roughly 5\%. However, modeling also the dynamics of the spike, which responds to incorporating feedback from the orbiting compact object, we find the dephasing effect is reduced to 0.07\%.  
As we saw in Fig.~\ref{fig:EDFbyUDMdensity}, previous calculations assuming a static DM spike overestimated the magnitude of energy loses compared to the binding energy in the DM spike by up to several orders of magnitude. In this case, we see that incorporating DM feedback is not a small correction, but instead reduces the size of the dephasing effect by roughly a factor of 100.

For a heavier central IMBH of $m_2 = 1.4 \times 10^4\,M_\odot$, the binding energy available in the DM spike is larger. As shown in Fig.~\ref{fig:EDFbyUDMdensity}, this available energy is on the same order as the work done by dynamical friction. This is reflected in the smaller difference between the results for the static and dynamic spikes. The dephasing would appear as a roughly 0.5\% effect if we assumed a static spike; the dephasing effect is reduced by a further 50\% once we incorporate dynamic feedback of the DM. For a spike with slope $\gamma_\mathrm{sp} = 7/3$, the dephasing effect still corresponds to a difference of around 5800 GW cycles.

We note that assuming a static halo, the size of the dephasing effect is smaller for a heavier IMBH because dynamical friction is subdominant to GW energy losses (for the initial separations we consider here). However, due to the tighter gravitational binding, the impact of allowing for a dynamic DM spike is smaller for a heavier IMBH. Thus, in the dynamic case, the dephasing effect is larger for a central BH of mass $m_1 = 1.4 \times 10^4\,M_\odot$ than for $m_1 = 1.4 \times 10^3\,M_\odot$. This suggests that a mass ratio $q = \mathcal{O}(10^{-4})$ is a promising target for detecting the effect of a DM spike on the gravitational waveform.

While the dephasing including halo feedback is still smaller than that predicted by \EdaEtAl, we expect that the qualitative conclusions of~\cite{Eda:2014kra} should still hold: namely, that the effects of the DM on the emitted GWs will allow properties of the DM distribution to be measured from the observed GWs by an interferometer like LISA.
We leave computation of how well LISA will be able to measure the properties of the DM spike to future work.

Finally, for a central IMBH of $m_1 = 1.4 \times 10^5\,M_\odot$, incorporating feedback appears to lead to a percent-level correction to the dephasing effect. Such percent-level corrections are important if we wish to model the IMRI waveform to high precision. However, the overall size of the dephasing effect is much smaller, and the difference in $\Delta N_\mathrm{cycles}$ between the static and dynamic case is typically smaller than our numerical accuracy of $\mathcal{O}(100)$ cycles. Even so, such a small difference is in line with our expectations from right panel of Fig.~\ref{fig:EDFbyUDMdensity}, which shows that the binding energy of the DM halo is typically larger than the work done by dynamical friction, due to the larger potential of the central IMBH. Further refinements to our numerical procedure will be required to determine the precise size of the dephasing effect in this case.

\begin{figure}[t!]
    \centering
    \includegraphics[width=\linewidth]{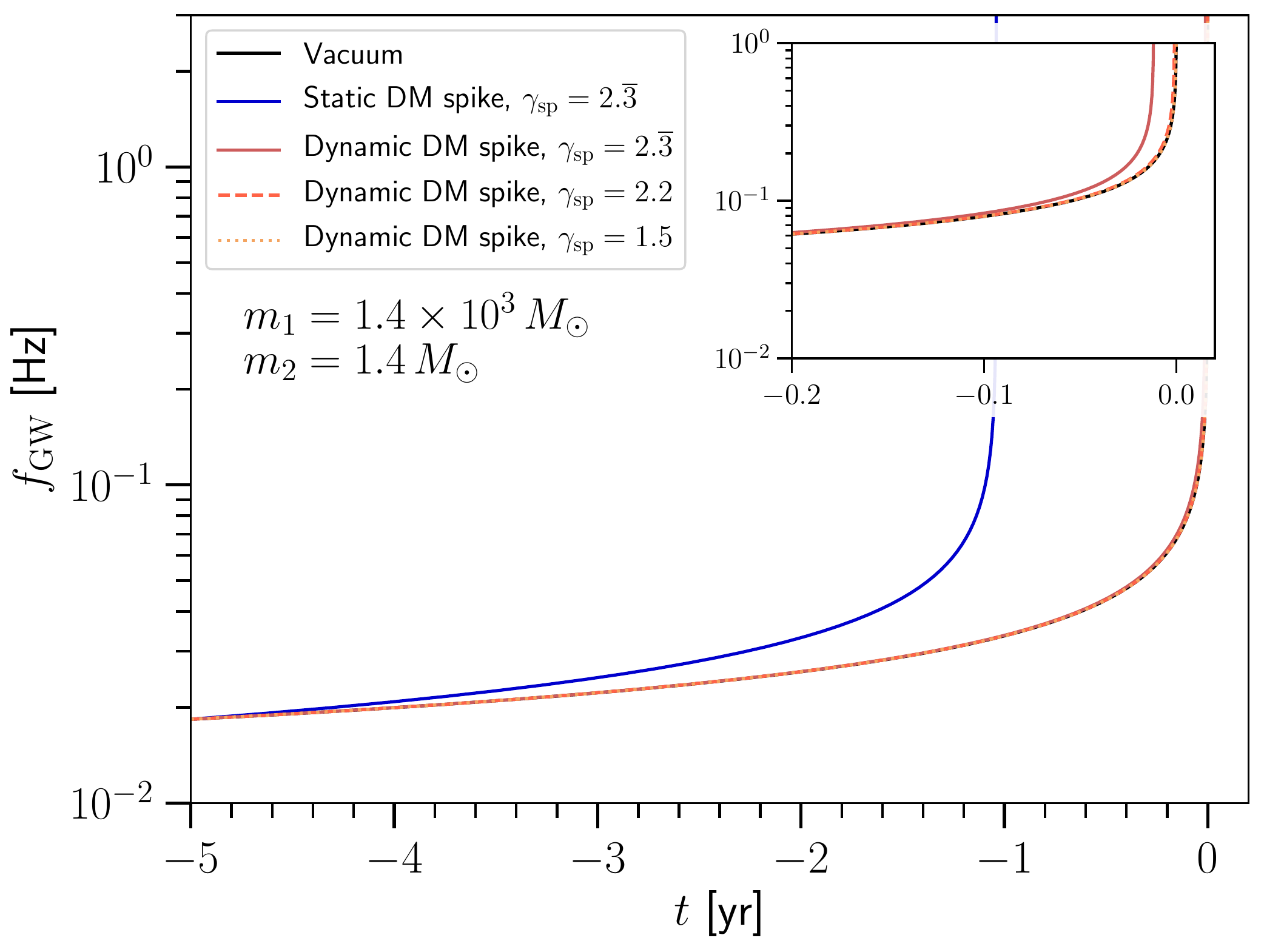}
     \includegraphics[width=\linewidth]{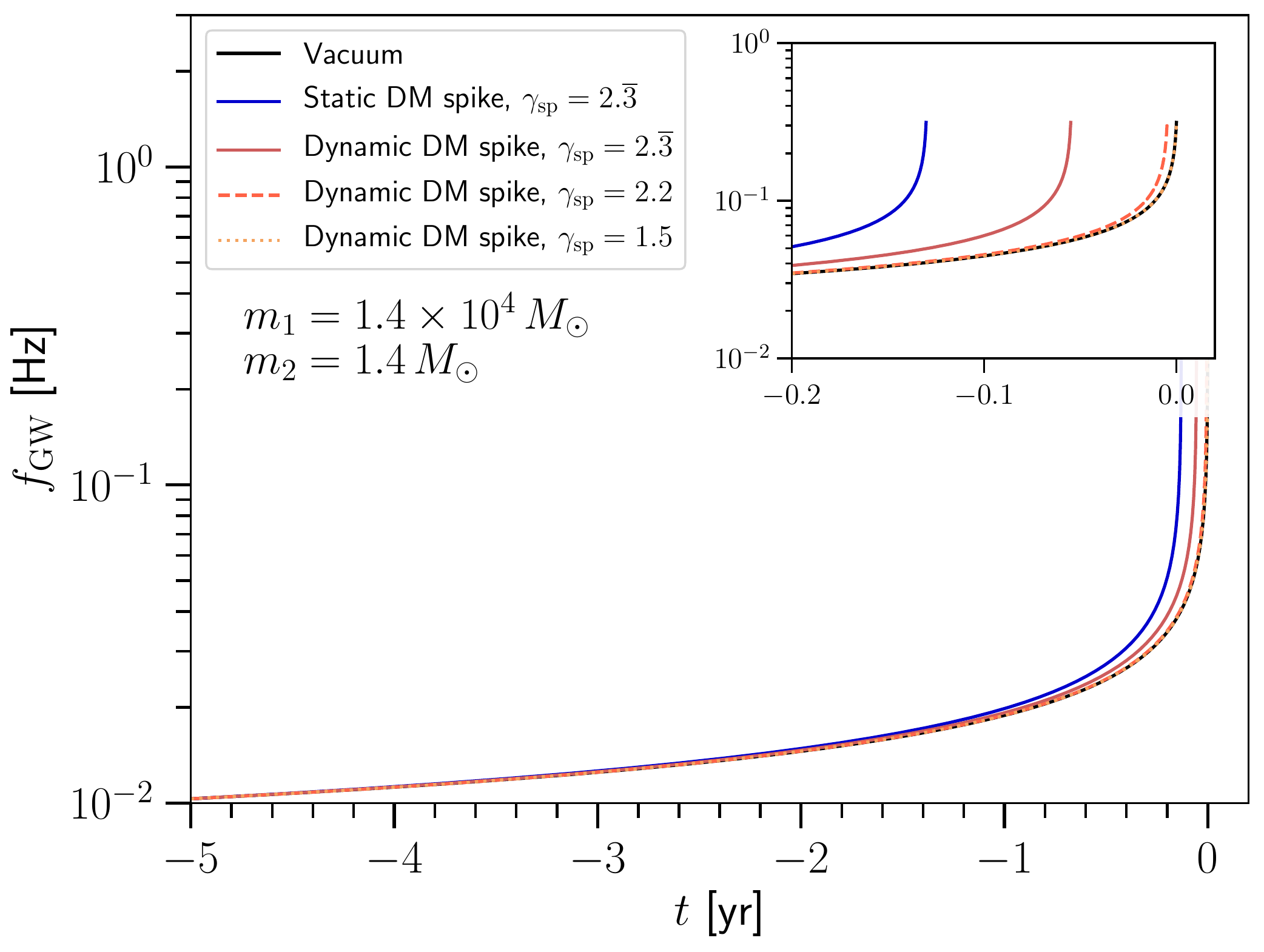}
    \caption{\textbf{Frequency evolution of the IMRI system.} Gravitational wave frequency of the binary as a function of time, starting approximately 5 years before the merger. The black curve shows the evolution in the absence of a DM spike, while the colored curves show the evolution for spikes with characteristic density $\rho_\mathrm{sp} = 226 \,M_\odot/\mathrm{pc}^3$ and different slopes $\gamma_\mathrm{sp}$. Note that $2.\overline{3} = 7/3$. \textbf{Top:} mass ratio $q = 10^{-3}$. \textbf{Bottom:} mass ratio $q = 10^{-4}$.}
    \label{fig:freq-evol}
\end{figure}


\begin{figure}[th!]
    \centering
    \includegraphics[width=\linewidth]{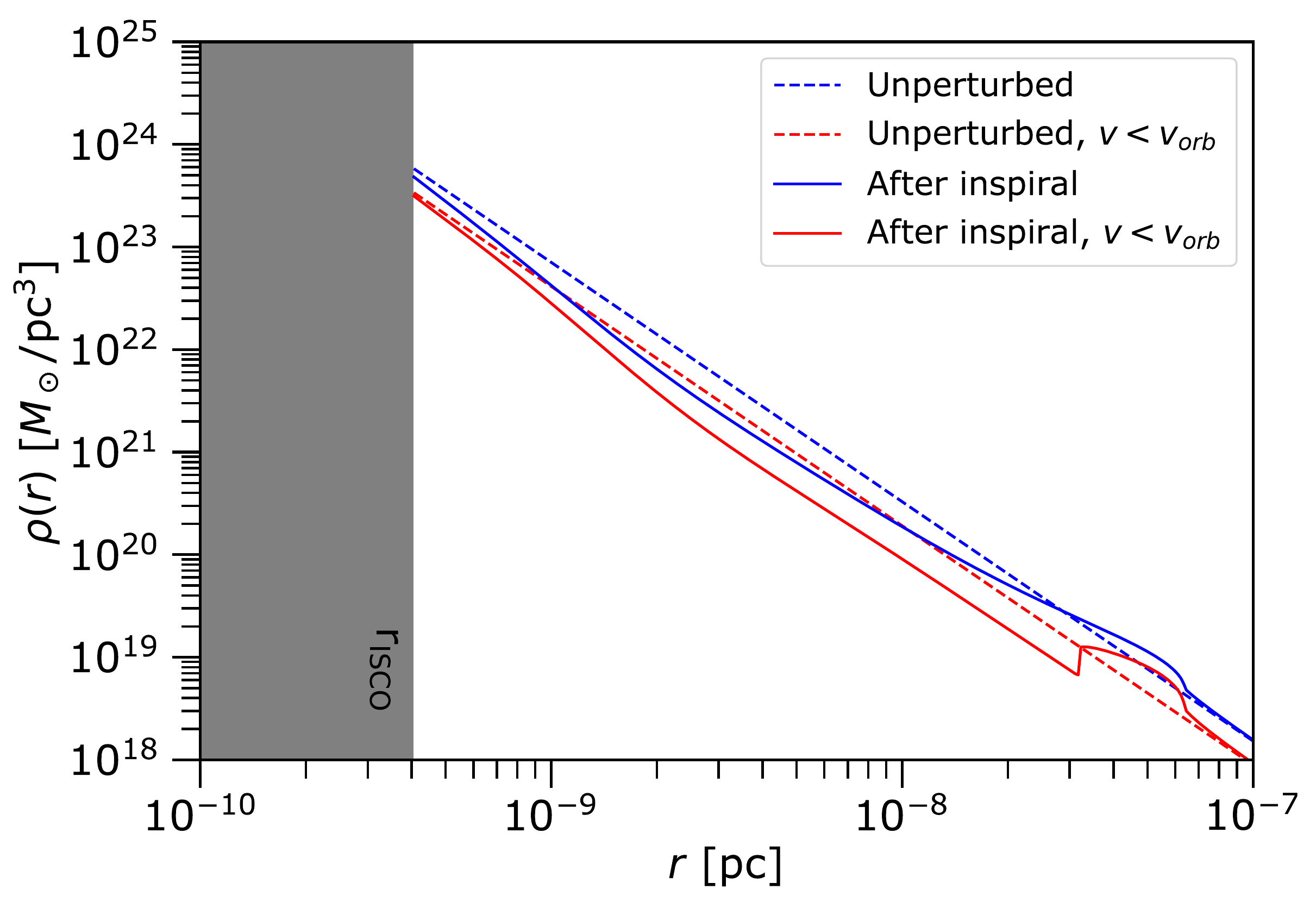}
    \caption{\textbf{Evolution of the DM mini-spike profile.} The solid lines refer to the system at the end of the inspiral, while the dashed lines correspond to the initial, unperturbed configuration. Blue lines: total density profile. Red lines: density profiles associated to the particles slower than the circular speed $v_\mathrm{orb}(r)$ for each $r$. The `bump' at $r \gtrsim 3 \times 10^{-8}\,\mathrm{pc}$ is an artifact of starting the compact object at this radius.}
    \label{fig:Profiles}
\end{figure}

As well as reducing the number of GW cycles, dynamical friction is also expected to shorten the inspiral time between two fixed frequencies \cite{Yue:2019ndw} and change the density profile of the DM mini-spike. In Fig.~\ref{fig:freq-evol}, we plot spectrograms, showing the frequency evolution of the GW signal with time, starting from a fixed initial frequency.  For a mass ratio $q= 10^{-3}$, the assumption of a static DM spike implies that a 5-year inspiral in vacuum would be shortened by more than 1 year in the presence of a DM spike with $\gamma_\mathrm{sp} = 7/3$.  However, our self-consistent  model substantially reduces the size of the effect, leading to an inspiral which is just 4 days shorter than the vacuum case. We also see that the inspiral time is very sensitive to the slope of the DM  distribution, rapidly becoming undetectable for a mild slope of $\gamma_\mathrm{sp} = 3/2$. For a mass ratio $q= 10^{-4}$, the impact of allowing for a dynamic spike is less extreme, though still gives an $\mathcal{O}(1)$ change. The inspiral is shorter by around 48 days in the static case, reduced to around 20 days in the dynamic case.  

Finally, we show in Fig.~\ref{fig:Profiles} the effect of the inspiral on the density profile of the DM mini-spike. Here, again we assume a central IMBH mass of $1.4 \times 10^3\,M_\odot$ and a fiducial spike with $\rho_\mathrm{sp} = 226 \,M_\odot/\mathrm{pc}^3$ and slope $\gamma_\mathrm{sp}= 7/3 = 2.333\ldots \equiv 2.\overline{3}$. We notice that, after the inspiral, the DM density at each radius is altered at most by a factor of $2$ with respect to the initial configuration (for a compact object that begins its inspiral at $r_2 \approx 3\times10^{-8}$~pc). 
This is because particles which scatter with the orbiting compact object are typically not completely unbound from the system but rather increase their average radius slightly. Thus, as the compact object inspirals, it depletes particles at its current radius, partially replenishing particles which were previously depleted further out. At small radii, the density profile is largerly unperturbed, as GW emission (and not dynamical friction) becomes the dominant energy loss mechanism here. While we have seen that feedback of the DM spike can have a dramatic impact on the dephasing signal, this does not mean that the spike is destroyed in the process. These results suggest that the DM overdensity may survive the inspiral with only a small amount of depletion overall. We expect also that the imprint of the inspiral on the DM spike will be too small  an effect  (and occur on too small an angular scale) to measure by other means (e.g.~dynamically or with multiwavelength electromagnetic observations).



\section{Discussion}
\label{sec:Discussion}

In this section, we discuss a number of caveats to the calculations we have performed. We suggest a number of avenues for improvements in the future as well as the prospects for detecting the effects of dark matter on the gravitational waveform.

\subsection{Halo relaxation}

Thus far, we have assumed that the DM halo is disrupted by the orbiting compact object and does not evolve further. We now consider processes which may \textit{replenish} the depleted halo. One possibility is that DM particles may diffuse in energy through small-angle scattering with each other, ultimately refilling the depleted regions. Following Refs.~\cite{1976ApJ...209..214B,1977ApJ...211..244L}, the relaxation time associated with this process scales as
\begin{align}
    \begin{split}
      t_\mathrm{relax} &\sim \frac{\sigma_v{}^3}{G^2 m_\mathrm{DM}^2 n_\mathrm{DM}}  \approx \frac{m_1^{3/2}}{G^{1/2}  \rho_\mathrm{DM} m_\mathrm{DM} r^{3/2}}\,,\\\
    \end{split}
\end{align}
where the DM velocity dispersion is approximately $\sigma_v^2 \approx  G m_1/r$. For a 100 GeV DM particle, we find $t_\mathrm{relax} \gtrsim 10^{70}$ years for the systems we consider here.

We may also worry about DM scattering with the compact object and losing energy, thereby replenishing the depleted regions of phase space. This process is only possible for DM particle moving more quickly than the orbiting object and would lead to a net ``cooling'' for these particles \cite[p.~582]{BinneyAndTremaine}. However, this process is suppressed with respect to the ``heating'' process we have considered here by a factor $m_\mathrm{DM}/m_2$ and can therefore be neglected. Without external perturbations, then, the disruption of the halo caused by the compact object should persist on timescales much longer than the inspiral time.

\subsection{Spherical Symmetry}

In Sec.~\ref{sec:HaloFeedback}, we relied on a description of the DM halo as spherically symmetric and isotropic. However, the binary is \textit{not} spherically symmetric so we eventually expect this description to break down. 

One possible issue is that the compact object scatters with particles in the DM spike only within a torus along its orbit (see Fig.~\ref{fig:Geometry}). Thus, energy is not injected into the halo in a spherically symmetric way. Of course, particles in the DM halo are not static; particles are on orbits which are (in general) inclined with respect to the orbital plane of the binary. Thus, energy injected in the plane of the orbit will be redistributed throughout the DM halo naturally through the dynamics of the system.

More concerning is the fact that the binary will inject angular momentum into the halo, just as it injects energy. On average, the scattered particles gain angular momentum and the halo begins to co-rotate with the binary. We can estimate how rapidly the halo is spun-up by calculating the typical change in the specific angular momentum of a DM particle $\left\langle  \Delta L \right\rangle$ each time it scatters. Comparing the torque on the compact object with the number of DM particles which scatter in a single orbit, we obtain:
\begin{equation}
    \left\langle  \Delta L \right\rangle = 4 \log \Lambda \, m_2 \sqrt{\frac{G r_2}{m_1}}\,.
\end{equation}
The maximum specific angular momentum at a given radius is achieved for circular orbits $L_\mathrm{max} = \sqrt{G m_1 r_2}$. We thus find that:
\begin{equation}
    \frac{\left\langle \Delta L \right\rangle}{L_\mathrm{max}} \approx \frac{4 \log \Lambda}{\Lambda^2} \approx 1\%\,,
\end{equation}
for a mass ratio of $q = 10^{-3}$. Thus, the spin of the DM halo increases only by a small amount with each scatter and $\mathcal{O}(100)$ interactions are required before a particle is expected to be on a circular orbit and co-rotating with the compact object. 

In a similar way, the typical change in the relative energy per unit mass of a DM particle can be calculated as:
\begin{equation}
    \frac{\left\langle \Delta \mathcal{E} \right\rangle}{\mathcal{E}} \approx -\frac{4 \log \Lambda}{\Lambda^2} \approx -1\%\,,
\end{equation}
where we have used the fact that the maximum energy for particles at radius $r_2$ is $\calE = G m_1/r_2$. 
Thus, by the time a particle has scattered enough to be spun up, it will have gained enough energy to become unbound. We therefore expect that the halo will not gain a substantial net angular momentum during the inspiral.

It is also possible to compute the amount of angular momentum radiated through dynamical friction for a static halo analogously to what was done in the calculations of energy dissipated through dynamical friction in Sec.~\ref{subsec:EnergyDissipate}.
Using the fact that for binaries in quasicircular orbits the angular momentum radiated satisfies $\mathrm{d}E_\mathrm{orb}/\mathrm{d}t = \Omega_\mathrm{orb} \mathrm{d}J_\mathrm{orb}/\mathrm{d}t$,  it is possible to show that the angular momentum dissipated through dynamical friction satisfies a relation analogous to Eq.~\eqref{eq:DeltaEDFhyp}: namely, it can be written as the change in  $\mu \sqrt{GMr_2}$ times a hypergeometric function (where the hypergeometric function for positive $r_2$ is again a number between zero and one). 
Thus the maximum amount of angular momentum dissipated via dynamical friction would go as $\mu \sqrt{GMr_\rmsp}$.
Because the angular momentum for the halo is assumed to be zero initially, it is not possible to compare the angular momentum dissipated to the amount of angular momentum in the halo, in analogy to the ratios of energy discussed in Sec.~\ref{subsec:EnergyRatio}.
Consider instead a simple toy model of a DM spike with a large angular momentum, in which each spherical shell of DM is rigidly rotating at the Keplerian orbital frequency. 
A straightforward calculation of the angular momentum of this spike shows that it would scale as $m_1 \sqrt{G M r_\rmsp}$.
Thus, the ratio of the angular momentum dissipated to the angular momentum of this rotating distribution goes as $\mu / m_1 \approx q$, which is small for the binaries that we have considered.
Because this ratio is small for the static halo, it should be smaller for the dynamic halo, because less energy (and thus also angular momentum) is radiated.

We note also that if more DM particles \textit{are} co-rotating, the size of the dynamical friction effect should increase. The relative velocity of encounters with the compact object will decrease, enhancing the drag force on the compact object, as described in Eq.~\eqref{eq:DFdissipation}. Thus, our approach may be seen as a conservative estimate of the size of the dephasing effect.

Ultimately, to obtain high precision waveforms, it will be necessary to follow both the energy \textit{and} angular momentum of DM particles in the halo. However, we expect the results we present here to be conservative, with corrections due to angular momentum injection being higher order. We defer this more detailed analysis to future work.

\subsection{Relativistic and other corrections to the binary}

Our focus in this paper was to understand the effects of jointly evolving the binary and the DM spike on the emitted GWs (and we found the effect can be substantial).
We made a number of simplifying approximations in modeling the orbital dynamics of the binary and the DM spike.
Because we made the same types of assumptions for the orbital dynamics with and without DM spikes, this allowed us to obtain a self-consistent estimate of the impact of an evolving DM spike on the GWs within the context of our assumptions.
However, because the detection of IMRIs with LISA using matched filtering usually requires gravitational waveform templates that are accurate to within a few orbital cycles of the binary, the orbital dynamics that we computed in this paper will likely not be sufficiently accurate to use for GW data analysis.
We now comment on the types of effects and calculations that we expect need to be added to make the gravitational waveforms more suitable for data analysis.

Most notably, we restricted our calculations throughout this work to a Newtonian description of the orbital dynamics of the binary and the DM halo. 
For the system with $q=10^{-3}$, the initial orbital velocity is given roughly by $(v/c)^2 \sim 0.01$, so post-Newtonian (PN) effects will produce a roughly 1\% error.
Because there are of order $10^6$ GW cycles during a five-year inspiral, these 1\% errors can lead to inaccuracies of order $10^4$ GW cycles.
While this error is greater than the dephasing shown in Table~\ref{tab:Ncycles}, this error will not contaminate our results for the following reasons: (i) these leading PN corrections here are corrections to the conservative dynamics, but the effect of dynamical friction is a dissipative effect, which will allow these effects to be distinguished; (ii) the dephasing signal occurs predominantly when the separation of the binary is large and when PN effects are small; and (iii) the dynamical friction corresponds to a \textit{negative} PN-order effect for quasicircular orbits, so it will not be confused with standard PN effects.\footnote{For the shell model, the effect is a $\gamma_\rmsp-3$ PN-order effect, whereas for a static halo, it is a $\gamma_\rmsp-11/2$ PN-order effect.
Because Fig.~\ref{fig:deltaN-cycles} showed that for the dynamic case, the power law of the effect is closer to the shell model, the PN-order will be closer to a $\gamma_\rmsp-3$ effect, though it will not be precisely a fixed PN order.
For $\gamma_\rmsp$ close to two, the effect might be mistaken for the effects of dipole radiation that appear in certain modified-gravity theories (see, e.g., the review in~\cite{Yunes:2013dva}).}

A more complete description of the dynamics of the system will be developed in future work. 
There we plan to incorporate a relativistic description of the orbital dynamics and distribution of dark matter.
We also intend to more carefully understand the effects of assuming the barycenter and the IMBH are collocated.
Finally, we will incorporate (and revise) the effects of accretion of DM when the small compact object is a black hole rather than a neutron star that were discussed in~\cite{Yue:2017iwc}.
Attempting to incorporate these effects goes beyond the scope of this initial work.

\subsection{Detection prospects}

For concreteness, we have focused on the final 5 years of the inspiral, having in mind a 5-year LISA mission. 
We chose the final 5 years of inspiral, because the amplitude of the GWs will be largest during this last stage of the inspiral, which will typically imply that the system would have the largest signal-to-noise ratio (though the precise signal-to-noise will depend upon the details of LISA's noise curve, the mass of the system, and the initial orbital frequency of the binary when the LISA mission begins).
Of course, there is no guarantee that the merger itself will occur during the LISA observation period (and because the binary spends more time orbiting at larger radii, it is likely that there will be more binaries at earlier stages in their inspiral). 
If the system is observed at an earlier time, further from the merger, the signal-to-noise ratio and the size of the dephasing would be different.

Determining the specific parameters of binaries and the stage in their orbital evolution for which the dephasing effect is most likely to be measured is an interesting, but more complex question, that we plan to consider in future work. We also postpone to a future analysis a discussion about the possibility that the effect considered in this work could be misinterpreted in the context of an actual ``real-world'' data analysis, and may lead to a biased estimation of the orbital parameters. For instance,  a larger mass of the central object (hence, a larger GW reaction force) could partially mimic the dynamic friction effect considered here, although the the dephasing due to friction is typically accumulated at larger radii.

In addition, in order to assess the prospects for detection, we must explore in detail how many such systems we expect to observe and with what properties. 
It is estimated that LISA will detect IMRIs at a rate of $\mathcal{R} \sim 3-10 \,\mathrm{Gpc^{-3}\,yr^{-1}}$~\cite{Fragione:2017blf}. However, only a fraction of these will be embedded in a DM spike. Very dense spikes are expected to form only at the centers of DM halos, around adiabatically growing BHs \cite{Ullio:2001fb}. In addition, spikes may be disrupted by mergers and other dynamical processes~\cite{Wanders:2014xia}. The presence of baryons may also affect the formation of the spike~\cite{Ullio:2001fb}, though there are a number of scenarios in which we do not expect these systems to be baryon-dominated (including direct-collapse IMBHs~\cite{Zhao:2005zr,Bertone:2006nq} and primordial IMBHs~\cite{Kohri:2014lza,Eroshenko:2016yve,Boucenna:2017ghj}). In any case, we emphasise that the formalism we have developed here for modeling the dephasing does not require a `pristine' spike; indeed, our method applies equally well to partially disrupted spikes. Taking all these effects into consideration will be important for understanding the likelihood that LISA will be able to detect such systems during its time of operation.

Clearly, a more exhaustive exploration of the parameter space is warranted, taking into account the population properties of IMRI systems, in order to assess detectability of the inspiral signal and associated dephasing. These topics will be addressed in follow-up work.

\section{Conclusions}
\label{sec:Conclusions}

Dark matter overdensities around intermediate mass black holes inevitably modify the dynamics of inspiraling compact objects, and 
could potentially be detected through their impact on the gravitational waveform produced by the binary inspiral. 

We have demonstrated that previous analyses have largely overestimated the dephasing induced by the dynamical friction experienced by the compact object ploughing through the dense dark matter spike. Those studies relied in fact on the simplifying assumption of a static dark-matter distribution, whereas we have shown here that there is an efficient transfer of energy from the binary to the dark-matter spike. The energy dissipated by dynamical friction can in fact be much larger than the binding energy in the DM distribution.  

Guided by $N$-body simulations, we have then introduced a prescription to update the dark-matter phase space density as the binary evolves. Dynamical friction in general speeds up the inspiral,  reducing the number of GW cycles which would be observed by experiments such as LISA. Compared to the case of a static spike, our prescription leads to a depletion of the DM density at the orbital radius, which in turn reduces the size of the dynamical friction force and thus slows the inspiral. This has dramatic consequences for the impact of the DM on the emitted GWs, and the interpretation of the signal.

For a central IMBH with $m_1 = 1.4 \times 10^3 \,M_\odot$ and orbiting compact object with $m_2 = 1.4\,M_\odot$, assuming a static DM spike with slope $\gamma_\mathrm{sp} = 7/3$, leads to a 5\% difference in the number of cycles with respect to the vacuum case. When the dynamical evolution of the spike is taken into account according to our prescription, we find that the difference is reduced by a factor of $\sim 100$, to 0.07\%. The effect tends to be smaller for higher mass ratios, as the DM spike is more tightly bound and less easily disrupted. For a heavier central IMBH of $m_2 = 1.4 \times 10^4\,M_\odot$, our prescription leads only to a 50\% difference in dephasing, with respect to the static case. The effect however still corresponds to 5800 GW cycles, which should be observable and distinguishable by LISA.

Dynamical friction significantly shortens the inspiral time. For a mass ratio $q=10^{-3}$, a 5-year inspiral in vacuum would be shortened by more than 1 year in the presence of a static DM spike with $\gamma_\mathrm{sp} = 7/3$.  We have however shown that incorporating the feedback on the dark-matter distribution leads to a difference in inspiral time with respect to the vacuum case of only 4 days. We also found that the dephasing effect is very sensitive to the slope  of  the  DM  distribution,  rapidly  becoming less than one gravitational-wave cycle for a mild slope of $\gamma_\mathrm{sp} = 3/2$.

In future work, we will focus on the observational implications of the dynamical dark-matter spike for the LISA mission.
This will include estimates of the rate of intermediate and extreme mass-ratio inspirals with dark-matter spikes, studies of the detection prospects for these systems, and assessments of how well the properties of the dark-matter spike can be inferred from the gravitational-waves measured by LISA.
We anticipate that these systems will be detectable and that they could provide information about the nature of dark matter.

\acknowledgments

The authors would like to thank Jonathan Baird, Priscilla Canizares, Adam Coogan, Tom Edwards, Tanja Hinderer and Samaya Nissanke for helpful discussions about this work. B.J.K.~also thanks Rebekka Bieri and Jonathan Coles for useful guidance about $N$-body simulations.

D.A.N.\ acknowledges the support of the Netherlands Organization for Scientific Research through the NWO VIDI Grant No.~639.042.612-Nissanke.
D.G.\ has received financial support through the Postdoctoral Junior Leader Fellowship Programme from la Caixa Banking Foundation (grant n.~LCF/BQ/LI18/11630014).
DG was also supported by the Spanish Agencia Estatal de Investigaci\'{o}n through the grants PGC2018-095161-B-I00, IFT Centro de Excelencia Severo Ochoa SEV-2016-0597, and Red Consolider MultiDark FPA2017-90566-REDC.

This work was carried out on the Dutch national e-infrastructure with the support of SURF Cooperative. Finally, we acknowledge the use of the Python scientific computing packages NumPy \cite{numpy} and SciPy \cite{scipy}, as well as the graphics environment Matplotlib \cite{Hunter:2007}.


\appendix

\section{A heuristic model based on ejecting spherical shells of dark matter}
\label{sec:ShellModel}

In this subsection, we introduce a prescription to evolve a compact binary with DM between separations $r_\rmi$ and $r_\rmf$ such that the total energy input into the DM distribution is equal to the binding energy of the spherical shell of DM between $r_\rmi$ and $r_\rmf$.
We implement this procedure as follows.
Instead of equating the rate of energy dissipation by GWs in Eq.~\eqref{eq:GWdissipation} to be equal to minus the rate of change of the orbital energy, we set the GW dissipation equal to the orbital energy minus the energy of a shell of DM of width $\mathrm{d}r_2$ at the radius $r_2$ of the circular orbit.
Thus, we write
\begin{equation} \label{eq:EbalanceShell}
    \frac{\mathrm{d}E_\rmGW}{\mathrm{d}t} = \frac{\mathrm{d}r_2}{\mathrm{d}t}\left(\frac{\mathrm{d}E_{\mathrm{orb}}}{\mathrm{d}r_2} - \frac{\mathrm{d}U_\rmsh}{\mathrm{d}r_2} \right) \, .
\end{equation}
In Eq.~\eqref{eq:EbalanceShell} the three derivatives of energies that appear can be obtained from Eqs.~\eqref{eq:Ushell},~\eqref{eq:Eorbit}, and~\eqref{eq:GWdissipation}, thereby leaving $\mathrm{d}r_2/\mathrm{d}t$ as the one unknown quantity.
Because the quantity multiplying $\mathrm dr_2/\mathrm dt$ in Eq.~\eqref{eq:EbalanceShell} is smaller than $\mathrm{d}E_\mathrm{orb}/\mathrm{d}r_2$, the system will inspiral more rapidly than it will in vacuum.
Consequently, the number of orbital (and GW) cycles that the binary undergoes when inspiraling between two radii will be smaller.
We will compute analytical expressions the number of cycles as a function of frequency for this model in Appendix~\ref{app:CyclesAnalytic}.

This model is heuristic in the sense that it assumes that all the binding energy in the dark-matter distribution around the black hole will be dissipated through the scatterings that induce dynamical friction on the small compact object.
It aims to provide a conservative, though still rough, upper limit on the size of the dephasing effect $\Delta N_\mathrm{cycles}$ that is likely to occur.
The results in Fig.~\ref{fig:deltaN-cycles} show that it captures some of the qualitative features of the dephasing effect of the dynamical halo feedback model, when feedback is significant, and that it does provide an upper bound on the magnitude of the effect.

\section{Analytical expressions for the number of gravitational-wave cycles}
\label{app:CyclesAnalytic}

In Sec.~\ref{sec:Results}, we quantified the size of the dephasing effect by computing the difference in the number of GW cycles between two frequencies or over a fixed amount of time.
Here we provide analytical expressions for the number of GW cycles between two frequencies in vacuum, for a static DM distribution, and for the shell model in Appendix~\ref{sec:ShellModel}.

To compute the number of GW cycles, we combine a number of results.
First, we take the expression for the number in cycles Eq.~\eqref{eq:NcyclesTime} and rewrite it as a function of the GW frequency as
\begin{equation}
    N_\mathrm{cycles}^\mathrm{vac} = \int^{f_\mathrm{GW,f}}_{f_\mathrm{GW,i}} f_{GW} \frac{\mathrm{d}t}{\mathrm{d}f_\rmGW} \,\mathrm{d}f_\rmGW \, .
\end{equation}
Then we comput the derivative $dt/df_\rmGW$ by using Kepler's law for the orbital frequency, the fact that $f_\rmGW = \Omega_\mathrm{orb}/\pi$, the expressions for the derivative $dr_2/dt$ [we will consider the three different cases given by Eq.~\eqref{eq:r_dot}, with and without dark matter, and Eq.~\eqref{eq:EbalanceShell}], and the chain rule.
In the simplest case, in vacuum, the computation gives the familiar result
\begin{equation} \label{eq:NcyclesVacuum}
    N_\mathrm{cycles} = \frac 1\pi \left. \left( \frac{8\pi G \mathcal M_c f}{c^3} \right)^{-5/3} \right|_{f_\mathrm{GW,i}}^{f_\mathrm{GW,f}} \, .
\end{equation}

For the static DM halo, the number of cycles is given by
\begin{equation}
\begin{split}
 & N_\mathrm{cycles}^\mathrm{DM}(f_\mathrm{GW,f}, f_\mathrm{GW,i}) = \Bigg[\frac 1\pi \left( \frac{8\pi G \mathcal M_c f}{c^3}\right)^{-5/3} \times \\
 & {}_2F_1\left(1, \frac{5}{11-2\gamma_\rmsp}, \frac{16-2\gamma_\rmsp}{11-2\gamma_\rmsp}; -c_f f^{-11/3+2\gamma_\rmsp/3} \right)\Bigg]\Bigg|^{f_\mathrm{GW,f}}_{f_\mathrm{GW,i}}
 \, .
\end{split}
\end{equation}
The coefficient $c_f$ is defined by
\begin{equation} \label{eq:NcyclesStaticDM}
     c_f = \frac{5 G c^5 q \rho_\rmsp r_\rmsp^{\gamma_\rmsp} \log\Lambda}{(G\mathcal M_c)^{5/3} (G M)^{\gamma_\rmsp/3} \pi^{(8-2\gamma_\rmsp)/3}}
\end{equation}
The hypergeometric function is a number between zero and one for positive frequencies $f_\rmGW$.
Like the result for the energy dissipated in Eq.~\eqref{eq:DeltaEDFhyp}, the result including the DM spike can be written as the difference of two fractions of the vacuum value at the relevant frequencies.

Finally, we can compute the number of cycles for the shell model of Appendix~\ref{sec:ShellModel}.
A similar calculation shows that
\begin{equation} \label{eq:NcyclesShell}
    N_\mathrm{cycles}^\mathrm{sh} = \frac 1\pi \left. \left( \frac{8\pi G \mathcal M_c f}{c^3} \right)^{-5/3} [1 - c_{\rmsh}(f)]\right|_{f_\mathrm{GW,i}}^{f_\mathrm{GW,f}} \, .
\end{equation}
where
\begin{equation}
    c_{\rmsh}(f) = \frac{40 \pi \rho_\rmsp r_\rmsp^{\gamma_\rmsp} }{(11-2\gamma_\rmsp)m_2} \left[ \frac{G M}{(\pi f)^2} \right]^{1-\gamma_\rmsp/3} \, .
\end{equation}
The term in square brackets is just $r_2^{3-\gamma_\rmsp}$, from which one can see that it has the form of a negative $3-\gamma_\rmsp$ PN-order effect.
Equations~\eqref{eq:NcyclesVacuum},~\eqref{eq:NcyclesStaticDM}, and~\eqref{eq:NcyclesShell} were used in Fig.~\ref{fig:deltaN-cycles}.

\section{$N$-body simulations}
\label{app:NbodyDetails}

Here, we provide more technical details about the $N$-body simulations described in Sec.~\ref{sec:Nbody}. We use the publicly available  \textsc{Gadget}-2 code \cite{Springel:2000yr,Springel:2005mi}, with minor modifications which we describe below. In order to specify initial conditions and read the \textsc{Gadget} snapshots in Python, we use \textsc{pyGadgetIC} \cite{pyGadgetIC} and \textsc{pyGadgetReader} \cite{2014ascl.soft11001T}.

We fix the softening length to be $\ell_\mathrm{soft} \approx 10^{-10}\,\mathrm{pc}$, approximately the Schwarzschild radius for a 1000 $M_\odot$ black hole. For the simulations using a central mass of 1000 $M_\odot$, we reduce the softening length by roughly a factor of 4. This enhances our sensitivity to the small dynamical friction effect, as described in the main text. We have modified \textsc{Gadget}-2 to allow for a different maximum timestep for the DM particles and the compact objects. We set the maximum timestep for DM particles to be comparable to the typical orbital period $\mathcal{O}(1000 \,\mathrm{s})$, while the timestep for the orbiting compact objects is set a factor of $10^{-6}$ smaller. This allows us to trace the binary separation with sufficient precision (as illustrated in Fig.~\ref{fig:separation}). A summary of the parameters used in the simulations is given in Tab.~\ref{tab:Gadget}.

\begin{table}[t]
\caption{\textbf{Summary of \textsc{Gadget-2} parameters.} The parameter \texttt{ErrTolForceAcc} controls the accuracy of force calculations, while \texttt{ErrTolIntAccuracy} determines the error in the time integration. We specify the softening lengths $\ell_\mathrm{soft}$, for which we use a slightly smaller value for simulations with $m_1$. Each simulation contains $2^{15} \approx 33000$ DM particles.}
\begin{ruledtabular}
\begin{tabular}{llll}
\texttt{ErrTolForceAcc} & \multicolumn{3}{l}{$10^{-5}$}\\
\texttt{ErrTolIntAccuracy} & \multicolumn{3}{l}{$10^{-3}$}\\
\texttt{MaxTimestep} (BH) [s] &  \multicolumn{3}{l}{$1.5 \times 10^{-3}$}\\
\texttt{MaxTimestep} (DM) [s] &  \multicolumn{3}{l}{$1.5 \times 10^{3}$}\\
\hline
$m_1 = $& $100 \,M_\odot$ & $300 \,M_\odot$ & $1000 \,M_\odot$ \\
\hline
$\ell_\mathrm{soft}$ [pc] & $10^{-10}$& $10^{-10}$ & $ 2.4 \times 10^{-11}$
\end{tabular}
\end{ruledtabular}
\label{tab:Gadget}
\end{table}

Our only other modification of \textsc{Gadget}-2 is to alter the hard-coded value of Newton's constant $G$. The release version of \textsc{Gadget}-2 uses a value $G = 6.672 \times 10^{-11} \,\mathrm{m}^3 \,\mathrm{kg}^{-1}\,\mathrm{s}^{-2}$. This value of a factor of $\sim 3 \times 10^{-4}$ smaller than the current recommended value for $G_N$ \cite{CODATA}. This discrepancy is comparable to the relative change in orbital radius which we are hoping to observe (see Fig.~\ref{fig:separation}). Thus, it was necessary to change the hard-coded value to match the current value used elsewhere in our analysis chain.

For the purposes of the simulations, we model the DM spike using a generalized NFW profile:
\begin{equation}
    \rho_\mathrm{DM} = \frac{\rho_\mathrm{sp}}{(r/r_\mathrm{sp})^{\gamma_\mathrm{sp}}(1+r/r_t)^\alpha} \, .
\end{equation}
We set $\alpha = 2$, so that the profile drops off rapidly above the truncation radius $r_t$. This produces an equilibrium configuration with the correct density profile in the inner region of interest (to within a few percent) while keeping the total mass of the spike computationally feasible. We set the truncation radius equal to
\begin{equation}
    r_t = 10^{-5} \, r_\mathrm{sp} \left( \frac{100 M_\odot}{m_1} \right)^{3/2}\,,
\end{equation}
which means that the total mass of the simulated spike is approximately the same for the different values of $m_1$ we consider. We use $N = 2^{15}$ DM particles in each simulation and have checked that the spike profile is stable on the timescales of our simulations.

Each binary is initialized on a circular orbit around the barycenter of the system. We follow the separation of the two compact objects as a function of time to directly measure the dynamical friction force. We perform simulations with at least 5 different random realizations of the DM spike in order to extract an estimate of the error. The results are reported in Figs.~\ref{fig:DF_panels} and \ref{fig:DF_mass}.

\section{Scattering probability}
\label{app:Scattering}

We wish to evaluate the probability that a particle with energy $\mathcal{E}$ scatters to an energy $\mathcal{E} + \Delta\mathcal{E}$. This is given in Eq.~\eqref{eq:Ps_3}, which we repeat here:
\begin{align}
    \begin{split}
        P_\mathcal{E}(\Delta\mathcal{E}) &= \frac{4\pi^2  r_2}{g(\calE)}\frac{b_{90}{}^2}{v_0^2} \left[ 1 + \frac{b_\star^2}{b_{90}{}^2}\right]^2 \times \\ 
        & \int
         \sqrt{2\left(\Psi(r[b_\star, \alpha])-\mathcal{E} \right)} \sin\left(\theta[b_\star, \alpha]\right)\,\mathrm{d}\alpha\,.
    \end{split}
\end{align}
We recall that $b_\star = b_\star(\Delta \calE)$ and that the integration is over values of $\alpha \in [0, \pi]$ such that $r[b_\star, \alpha] \in [r_\mathrm{cut}, r_\calE]$. It is useful to recall also that $\Psi(r) = G m_1/r$ and
\begin{align}
\begin{split}
    r &= \sqrt{r_2^2 + b_\star^2 + 2r_2 b_\star \cos\alpha }\,,\\
   \sin\theta &= \frac{r_2 + b_\star \cos\alpha}{\sqrt{r_2^2 + b_\star^2 + 2r_2 b_\star \cos\alpha }}\,.
\end{split}
\end{align}
Expanding to first order in $(b/r_2)$, we obtain:
\begin{align}
\begin{split}
    r &\approx r_2 + b_\star\cos\alpha + \mathcal{O}\left(b_\star^2\right)\\
    &\approx \frac{r_2}{1 - (b_\star/r_2) \cos\alpha}\,.
\end{split}
\end{align}
This in turn gives:
\begin{align}
\begin{split}
\sin\theta &\approx 1 + \mathcal{O}\left(b_\star^2\right)\,,\\
    \Psi(r) & \approx \Psi(r_2) \left(1 - (b_\star/r_2)\cos\alpha + \mathcal{O}\left(b_\star^2\right)\right)\,.
\end{split}
\end{align}
The integral over the angle $\alpha$ can then be written:
\begin{align}
\begin{split}
    &\int_{\alpha_1}^{\alpha_2} \sqrt{2\left(\Psi(r) - \calE\right)} \,\mathrm{d}\alpha = 2\sqrt{2\Psi(r_2)} \sqrt{1 - \frac{r_2}{r_\calE} + \frac{b_\star}{r_2}}  \\
    &\qquad\times \left[E\left(\frac{\pi - \alpha_1}{2},m\right) - E\left(\frac{\pi - \alpha_2}{2},m\right)\right]\,,
\end{split}
\end{align}
where $E(\varphi, m)$ is the incomplete elliptic integral of the second kind:
\begin{equation}
    E(\varphi, m) = \int_{0}^\varphi \sqrt{1 - m\sin^2\theta}\,\mathrm{d}\theta\,,
\end{equation}
and
\begin{equation}
    m = \frac{2 (b_\star/r_2)}{1 - \frac{r_2}{r_\calE} + \frac{b_\star}{r_2}}\,.
\end{equation}

The limits of integration are set by requiring $r\in [r_\mathrm{cut}, r_\calE]$ which gives, again to first order in $(b/r_\star)$:
\begin{align}
\begin{split}
    \alpha_1 &= \cos^{-1}\left\{\min\left((r_2  - r_2^2/r_\calE)/b_\star, 1\right)\right\}\,,\\
    \alpha_2 &= \cos^{-1}\left\{\max\left((r_2 - r_2^2/r_\mathrm{cut})/b_\star, -1\right)\right\}\,.
\end{split}
\end{align}

The scattering probability $P_\calE( \Delta\calE)$ can now be evaluated in terms of special functions.\footnote{On a technical note, the SciPy implementation of $E(\varphi, m)$ is valid only for $m \leq 1$. To extend to values of $m>1$, it is necessary to perform reciprocal modulus transformations; see \href{https://dlmf.nist.gov/19.7\#ii}{Eq.~(19.7.4)} in Ref.~\cite{NIST:DLMF}.} With this, there is only a single numerical integral (over $\Delta \calE$) to be performed to evaluate $\partial f/\partial t$ in Eq.~\eqref{eq:dfdt}.


\bibliography{references}

\end{document}